\documentclass[useAMS,usenatbib,usegraphicx]{mn2e}

\def\lya{\ifmmode {\mbox{Ly}\,\alpha}\else
             {Ly\,$\alpha$}\fi}
\def\ergs{\ifmmode {\rm\,erg\,s^{-1}}\else
             ${\rm\,erg\,s^{-1}}$\fi}
\def\tnj1338{\ifmmode {\mbox{TNJ}1338-1942}\else
             {TNJ1338}\fi}

\title[The environment of \lya\ blob]
{The environments of \lya\ blobs I: 
Wide-field \lya\ imaging of TN~J1338-1942, 
a powerful radio galaxy at $z\simeq 4.1$ associated with a giant \lya\ nebula
\thanks{Based on data collected at Subaru Telescope, which is operated 
by the National Astronomical Observatory of Japan. }}
\author[T. Saito et al.]{Tomoki Saito$^{1,2}$\thanks{E-mail:
tomoki.saito@ipmu.jp (TS), and yuichi.matsuda@nao.ac.jp (YM)}, 
Yuichi Matsuda$^{3,4}$, Cedric G. Lacey$^{5}$, Ian Smail$^{5}$, 
Alvaro Orsi$^{6,7}$, \newauthor 
Carlton M. Baugh$^{5}$, Akio K. Inoue$^8$, Ichi Tanaka$^9$, Toru Yamada$^{10}$, 
Kouji Ohta$^{11}$,
\newauthor Carlos De Breuck$^{12}$, Tadayuki Kodama$^{3,4}$, 
Yoshiaki Taniguchi$^{13}$
\\
$^{1}$Institute for Cosmic Ray Research, The University of Tokyo, 
5-1-5 Kashiwanoha, Kashiwa, Chiba 277-8583, Japan\\
$^{2}$Kavli Institute for the Physics and Mathematics of the Universe (WPI), 
Todai Institutes for Advanced Studies, The University of Tokyo,\\
5-1-5 Kashiwanoha, Kashiwa, Chiba 277-8583, Japan\\
$^{3}$National Astronomical Observatory of Japan, 2-21-1 Osawa, Mitaka, 
Tokyo 181-8588, Japan\\
$^{4}$The Graduate University for Advanced Studies (SOKENDAI), 
2-21-1 Osawa, Mitaka, Tokyo 181-0015, Japan\\
$^{5}$Institute for Computational Cosmology, 
Department of Physics, Durham University, South Road, 
Durham DH1 3LE, UK\\
$^{6}$Instituto de Astrofisica, Facultad de Fisica, 
Pontificia Universidad Cat\' olica, 
Av. Vicu\~ na Mackenna 4860, Santiago, Chile\\
$^7$Centro de Astro-Ingenieria, Pontificia Universidad Cat\' olica, Av. 
Vicu\~ na Mackenna 4860, Santiago, Chile\\
$^8$College of General Education, Osaka Sangyo University, 3-1-1, 
Nakagaito, Daito, Osaka 574-8530, Japan\\
$^9$Subaru Telescope, National Astronomical Observatory of Japan, 
650 North A'ohoku Place, Hilo, HI 96720, USA\\
$^{10}$Astronomical Institute, Tohoku University, Aramaki, Aoba, 
Sendai 980-8578, Japan\\
$^{11}$Department of Astronomy, Kyoto University, Kyoto 606-8502, Japan\\
$^{12}$European Southern Observatory, Karl Schwarzschild Stra\ss e 2, 
85748 Garching, Germany\\
$^{13}$Research Center for the Space and Cosmic Evolution, Ehime University, 
2-5 Bunkyo-cho, Matsuyama, Ehime 790-8577, Japan
}
\begin{document}

\date{Draft version 29 Nov. 2014}

\pagerange{\pageref{firstpage}--\pageref{lastpage}} \pubyear{2014}

\maketitle

\label{firstpage}

\begin{abstract}
We exploit wide-field \lya\ imaging with Subaru to probe the environment 
around TN~J1338--1942, a powerful radio galaxy with a $>100\, \rm kpc$ 
\lya\ halo at $z=4.11$. We used a sample of \lya\ emitters (LAEs) down to 
$\log(L_{\rm Ly\alpha} [\ergs])\sim 42.8$ to measure the galaxy density 
around \tnj1338, compared to a control sample from a blank field taken 
with the same instrument. We found that \tnj1338\ resides in a region 
with a peak overdensity of $\delta_{\rm LAE}=2.8\pm 0.5$ on scales of 
$8\, h^{-1}\rm Mpc$ (on the sky) and $112\, h^{-1}\rm Mpc$ (line of sight) 
in comoving coordinates. Adjacent to this overdensity, we found a strong 
underdensity where virtually no LAEs are detected. We used a semi-analytical 
model of LAEs derived from the Millennium Simulation to compare our results 
with theoretical predictions. While the theoretical density distribution 
is consistent with the blank field, overdense regions such as that around 
\tnj1338\ are very rare, with a number density of 
$6.4\times 10^{-8}\rm Mpc^{-3}$ (comoving), corresponding to the 
densest $< 0.4$ percentile at $z\simeq 4.1$. 
We also found that the \lya\ luminosity function in the \tnj1338\ field 
differs from that in the blank field: 
the number of bright LAEs ($\log(L_{\rm Ly\alpha}[\ergs]) \ga 43.3$) 
is enhanced, while the number of fainter LAEs is relatively suppressed. 
These results suggest that some powerful radio galaxies associated with 
\lya\ nebulae reside in extreme overdensities on 
$\sim 3$--$6\, \rm Mpc$ scales, 
where star-formation and AGN activity may be enhanced via frequent 
galaxy mergers or high rates of gas accretion from the surroundings. 
\end{abstract}


\section{Introduction}
Galaxies are thought to form through accretion of baryonic material 
(Pre-Galactic Medium, PGM) from their surrounding environment, 
which provides the cold gas necessary for the star-formation process 
to proceed. The classical picture of this accretion is that the PGM 
is immediately heated to the virial temperature of the dark haloes 
(typically $\sim 10^6$\,K) to form hot haloes \citep[e.g.][]{rees1977}. 
Recent theoretical studies, in contrast, predict that cold gas 
accretes in the form of cold flows (with temperatures of $\la 10^5$\,K) 
which penetrate the hot haloes surrounding the galaxies and which 
over cosmic time maintain the star-formation activity in galaxies
\citep[e.g.][]{keres2005,dekel2009}. At these low temperatures the 
gas in the PGM will predominantly cool via \lya\ emission. 
The star-formation episodes powered by this accretion in turn provide a
stellar feedback mechanisms which can then re-heat and eject material 
in the form of galactic-scale outflows 
\citep[galactic winds: e.g.][]{veilleux2005,steidel2010}, 
which subsequently interact with the ambient material in the 
Circum-Galactic Medium (CGM). 
Hence both the accretion and outflows mean that galaxies interact 
with their surrounding environments, potentially resulting in 
extended \lya\ emission either through cold accretion of the PGM 
\citep[e.g.][]{fardal2001,dijkstra2006,
dijkstra2009,faucher-giguere2010,goerdt2010}, 
or from galactic winds ejected into CGM 
\citep[e.g.][]{tenoriotagle1999,taniguchi2000,mori2006,dijkstra2012}. 
Such \lya\ nebulae may therefore reflect either the structure of the 
surrounding material, or gas circulation processes during the 
formation/evolution of these galaxies \citep{furlanetto2005}. 

The precise mechanisms responsible for forming \lya\ nebulae are still 
unclear. Some \lya\ nebulae (e.g., those in the SA\,22 protocluster at 
$z=3.1$) have been  extensively observed, and are thought to be driven 
by starburst events in massive galaxies forming galactic winds 
\citep[e.g.][]{ohyama2003,wilman2005,matsuda2006,
geach2005,geach2009,uchimoto2008,uchimoto2012}. 
Moreover, recent surveys targeting  \lya\ ``blobs'' have also suggested 
a possible connection between their properties (specifically morphologies) 
and the environments they reside in \citep{erb2011,matsuda2011,matsuda2012}. 
Such an environmental dependence is also qualitatively consistent 
with the observations of relatively isolated \lya\ nebulae, which may 
point to the possibility that these are accretion-driven 
\citep{nilsson2006,smith2007,saito2008}. 
However, to determine the true significance of environment in the formation 
of extended \lya\ emission haloes, more detailed and systematic observations 
with sufficient depth and area are essential. 

It has been long known that powerful radio galaxies at high redshifts are 
often associated with large, extended \lya\ nebulae \citep[e.g.][]
{chambers1990,rottgering1995,vanojik1996,vanojik1997,venemans2007,
villar-martin2007}. Although these \lya\ nebulae appear linked to 
the radio emission, their properties are similar to the \lya\ blobs 
described above. For example, the large velocity widths exceeding 
$\sim 1000\rm\, km\, s^{-1}$ \citep{vanojik1997} or absorption across 
the full extent of the \lya\ nebulae \citep{rottgering1995,vanojik1996} 
are also seen in the \lya\ nebulae in protoclusters 
\citep[e.g.][]{wilman2005,matsuda2006}. 
Furthermore, observational studies suggest that large fraction of 
very luminous \lya\ blobs ($\ga 10^{44}\ergs$) are likely to be 
driven by obscured AGNs \citep[e.g.][]{colbert2011,overzier2013}. 
There appears to be a size dependence of the \lya\ nebulae on the size of 
radio sources \citep{vanojik1997}, as well as a hint of environmental 
dependence on the size of the \lya\ nebulae \citep{venemans2007}. 
Even for those \lya\ blobs not directly associated with radio sources, 
these are sometimes found to be harbouring AGNs 
\citep[e.g.][]{basu-zych2004,geach2009}. 
These results suggest that \lya\ nebulae are common phenomena reflecting 
the interaction between massive galaxies and their surrounding 
environments, regardless of the presence of AGNs. 
It is thus quite essential to probe the environments of \lya\ 
nebulae with and without radio sources. 

We have conducted a systematic, wide-field imaging observations 
of the environments around known giant \lya\ nebulae, as traced by 
\lya\ emitters (LAEs). Here we present the results from the imaging 
of the first target, the powerful radio galaxy TN~J1338--1942 
(hereafter \tnj1338) located at $z=4.11$ \citep{debreuck1999}. 
This radio galaxy is known to have a \lya\ nebula extending up to 
$\sim 100\,\rm kpc$ with an asymmetric morphology \citep{venemans2002}. 
It has been suggested that this source is associated with a highly 
overdense region traced by LAEs and Lyman break galaxies, 
and thus represents an ancestor of present-day clusters, i.e., a 
{\em proto}cluster 
\citep{miley2004,venemans2002,venemans2007,intema2006,overzier2008}. 
Theoretical simulations also suggests that the overdensity associated 
with \tnj1338\ is likely to be  an ancestor of a present-day cluster 
\citep{chiang2013}. Multiwavelength observations have shown that 
this source has radio lobes with an extent of $\sim 70\,\rm kpc$. 
\citep{debreuck2004, venemans2007}. 
The extended \lya\ nebula around this source has signatures 
of star-formation induced by the radio jet, well outside the host galaxy 
at radii of $\sim 20$ kpc or more \citep{zirm2005}. 
Finally, {\it Chandra} X-ray observations indicate that there is weak 
extended emission ($\sim 30\,\rm kpc$) which may arise from inverse-compton 
scattering of cosmic microwave background or locally-produced far-infrared 
photons \citep{smail2013}. 

Together these observations suggest a number of processes responsible 
for intense interaction between the central radio galaxy and the 
surrounding environment. However, due to the lack of the wide-field 
imaging data, the surrounding environment of this radio galaxy is 
still not well characterised. The existing LAE survey data cover a small 
field (two $7'\times 7'$) making it hard to map any overdensity, 
and there is no matched blank field observations which means
that even the magnitude of the overdensity is uncertain.
(\citet{venemans2007} used the Large-Area Lyman Alpha (LALA) survey 
results \citep{rhoads2000,dawson2004} for the reference, 
which are taken with the different telescope and instrument, 
targeting a slightly different redshift). 
Accordingly the relationship between the structure hosting the radio 
galaxy (and the protocluster) and the wider surrounding environment 
cannot be probed with the existing data. 

The goal of our study is to better quantify the environment of the 
powerful radio galaxy \tnj1338. To this end we have obtained wide-field 
intermediate-band images to identify the LAEs lying at the redshift 
of the radio galaxy, and so quantify the environment using the LAE number 
density. The magnitude of the overdensity was determined from comparison 
with a control sample taken from a blank field, SXDS \citep{furusawa2008}, 
at the same redshift and with the same instrumental setup \citep{saito2006}. 
This enables us to probe ``{\em on what scale does a significant 
overdensity exist?}'', i.e., the amplitude of the overdensity and the 
spatial extent of the protocluster. We also compare our observations with 
predictions derived from combing the Millennium Simulation 
\citep{springel2005} with a semi-analytical model for LAEs \citep{orsi2008}. 
This allows us to estimate how rare is the  overdensity found around 
the radio galaxy, and more generally how well the simulation can 
reproduce such environments. This is the first truly panoramic 
($\sim 30' \times 30'$) systematic and quantitative study of the 
environment of a radio galaxy field at $z\sim 4$. 

The rest of this paper is organized as follows: We introduce the 
data and observational details in \S\ref{sect:data}. The  
procedure to select LAEs from these data are presented in \S\ref{sect:sample}. 
In \S\ref{sect:results}, we compare  the observed density field and
luminosity function of LAEs in the radio galaxy field and 
the blank field, and then compare the observations to the mock LAEs 
from the theoretical simulations. 
Throughout this paper, we use the standard $\Lambda$CDM cosmology with 
$\Omega_M=0.3, $$\Omega_\Lambda=0.7$, $H_0 = 100h = 70\rm km\, s^{-1}Mpc^{-1}$, 
unless otherwise noted. All the magnitudes are in AB system 
\citep{oke1974,fukugita1995}.

\section[]{The data}
\label{sect:data}

\begin{table}
\centering
\begin{minipage}{85mm}
  \caption{Summary of Observations }
  \label{tab:obs}
  \begin{tabular}{@{}ccccc@{}}
  \hline
 Filter & $\lambda_{\rm cent}$/$\Delta \lambda$ $^a$ & Exposure Time  & $m_{5\sigma}$ $^b$ & PSF $^c$\\
 & [\AA/\AA] & [sec] & [AB mag] & [$''$]\\
\hline
 {\it IA624} & 6226/302 & 28800 ($1800 \times 16$) & 26.94 & 0.74 \\
 $B$ & 4417/807 & 4800  ($1200 \times 4$) & 26.87 & 0.80 \\
 $R$ & 6533/1120 & 5940  ($540 \times 11$) & 26.92 & 0.55 \\
 $i'$ & 7969/1335 & 2700 ($300 \times 9$) & 26.56 & 0.64 \\
 $BR^d$ & 6226/1927 & -- & 26.90 & 0.74 \\
\hline
\end{tabular}
$^a$The central wavelength and FWHM of the filters.\\
$^b$The $5 \sigma$ limiting magnitudes within a $1.5''$ diameter aperture.\\
$^c$The median PSF size (FWHM).\\
$^d$The weighted mean of the $B$ and $R$-band images.\\
\end{minipage}
\end{table}

\subsection{Subaru imaging data}
\begin{figure}
\includegraphics[width=80mm]{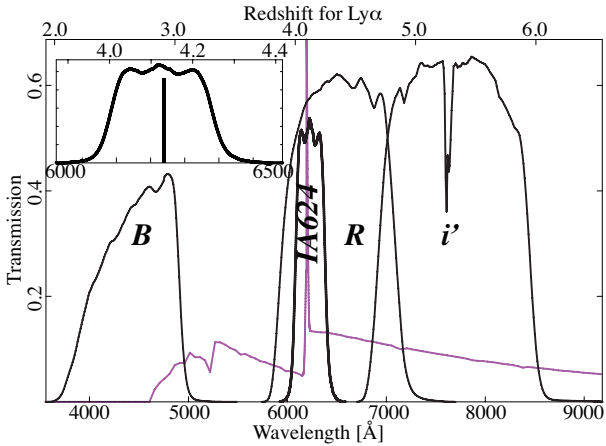}
\caption{The response curves of our filters. 
The transmission of the filters are plotted against the wavelength. 
The lower axis shows the observed wavelength, and the upper axis 
gives the corresponding redshift for \lya\ line. 
The thick black curve shows the intermediate-band filter ({\it IA624}) 
used to select sources with \lya\ emission. The two broadband filters 
({\it B} and {\it R}) are shown with thin lines. The thick magenta 
curve shows a model SED of a star-forming galaxy created with 
{\it GALAXEV} \citep{bc03} with a Gaussian-profile \lya\ emission line, 
redshifted to $z=4.1$. The box in the upper-left shows a close-up view 
of the {\it IA624} filter transmission, together with the redshift 
of \tnj1338\ indicated with a vertical thick bar. 
}
\label{fig:filters}
\end{figure}

We obtained 8.0 hours integration through the {\it IA624} 
intermediate-band filter centred at 
$(\alpha,\delta) = (13^h38^m13.9^s, -19^\circ42'29'')$ (J2000.0) 
on 2008 May 31 and June 01 (UT) with Suprime-Cam 
\citep{miyazaki2002} on the 8.2-m Subaru Telescope \citep{iye2004}, 
under the proposal ID S08A-072 (PI: Y.\ Matsuda). 
For the continuum correction we obtained 
archival broad-band imaging in the $B$ and $R$-bands 
taken by \citet{intema2006} from the archive system 
SMOKA \citep{baba2002}. 
The archival $i'$-band data by the same observers were also obtained 
to check the contamination using the continuum colour. 
Details of the observations are summarized in Table~\ref{tab:obs}. 
Suprime-Cam has a pixel scale of $0''.202$ and a field of view of 
$34' \times 27'$. The intermediate-band filter, {\it IA624}, has 
a central wavelength of 6226\AA\ and bandwidth of 302\AA\ (FWHM), 
which corresponds to the redshift range for \lya\ at $z=3.996$--$4.245$ 
\citep[{\it R23 IA filter system}: ][]{hayashino2000,taniguchi2004}. 
Fig.~\ref{fig:filters} shows the transmission curves of the {\it IA624}, 
$B$ and $R$-band filters, and the observed-frame wavelength 
of the \lya\ line at the redshift of \tnj1338\ ($z=4.11$). 

The raw data were reduced with {\sc sdfred20080620} 
\citep{yagi2002, ouchi2004} and {\sc iraf}. 
We flat-fielded using the median sky image after masking sources. 
We then subtracted the sky background adopting a mesh size 
of 64 pixels ($13''$) before combining the images. 
Photometric calibration was obtained from the spectroscopic standard 
stars, PG1323$-$086, and PG1708+602 \citep{massey1988, stone1996}. 
The magnitudes were corrected for Galactic extinction of $E(B-V)=0.10$ 
mag \citep{schlegel1998}. The variation of the extinction in this field 
is sufficiently small ($\pm 0.01$ mag from peak to peak) 
that it does not affect our results. 

The combined images were aligned and smoothed with Gaussian kernels to 
match their PSF to a FWHM of $0''.74$. The PSF sizes for some exposures 
in the $B$ band were not as good as the other bands (the median PSF size 
was $0.8''$), so that we removed these bad-seeing data, using only 
the four best frames for our analysis. The total size of the field 
analysed here is $32' \times 24'$ after the removal of low-S/N regions 
near the edges of the images. We also masked out the haloes of the 
bright stars within the field, resulting in the effective area of 
689 arcmin$^2$ in total. This corresponds to a comoving volume of 
$1.7 \times 10^5 h^{-3}\rm Mpc^3$ at $z=4.1$, covering a radial comoving 
distance of $112.4\, h^{-1}\rm Mpc$ for sources with \lya\ emission 
lying within the wavelength coverage of the {\it IA624} filter. 

The blank-field data used for the control sample were taken as part of the 
Subaru/{\em XMM-Newton} Deep Survey (SXDS) project \citep{furusawa2008} 
and a subsequent intermediate-band survey \citep{saito2006}. 
The SXDS field consists of five pointings, centred at 
$(\alpha, \delta) =  (02^h18^m00.0^s, -5^\circ00'00.0'')$ (J2000). 
The intermediate-band (including {\it IA624}) data were taken only 
in the south field (SXDS-S) centred at 
$(\alpha, \delta) =  (02^h18^m00.0^s, -5^\circ25'00.0'')$ (J2000). 
The data were taken with the same instrument on the same telescope, 
and reduced in the same manner with the same software as our \tnj1338\ 
observations. 
The PSF was matched to $0''.78$ (FWHM), and the $5\sigma$ limiting 
magnitudes are 26.60, 27.42 and 27.91 for the {\it IA624}, $B$ and $R$, 
respectively. After masking out the regions near the edges and bright 
stars, the total area coverage was 691 arcmin$^2$, which is almost the 
same as the \tnj1338\ field.

\subsection{Mock catalogue of \lya\ emitters}
In order to compare the observations with the predictions from a cosmological 
$N$-body simulation \citep[Millennium Simulation:][]{springel2005}, 
we exploited a LAE catalogue generated with a semi-analytical model of 
galaxy formation \citep[GALFORM][]{cole2000,ledelliou2005,ledelliou2006}. 
This catalogue contains $\sim 10^6$ LAEs with a wide range of 
luminosity and the \lya\ equivalent widths (EWs). 
The details of this catalogue are described in \citet{orsi2008}. 
The LAEs in this catalogue were generated by placing model galaxies 
into dark haloes of masses above the threshold value appropriate for 
the simulation's mass resolution. The star-formation history for each 
halo was calculated by using a Monte-Carlo merger tree. The IMF was 
assumed to be top heavy for those stars formed during any starbursts, 
$dN/d\ln(m) \propto m^{-x}$ with $x=0$, with a standard solar 
neighbourhood IMF \citep{kennicut1983} for those stars which quiescently 
form in discs, i.e.\ $x=0.4$ for $m<1M_\odot$ and $x=1.5$ for $m>1M_\odot$. 
Both IMFs covers the stellar-mass range $0.15\, M_\odot <m<125\, M_\odot$. 
The \lya\ line luminosity is then calculated from the number of
ionising photons emitted by the corresponding stellar population. 

\citet{orsi2008} compared the properties of model LAEs from their 
simulation with existing LAE surveys and showed that both the luminosity 
function (LF) and the clustering properties are broadly consistent with 
the observations. The number density of model LAEs in the luminosity 
range of $\sim 10^{42.5}$--$10^{43}\ergs$ are slightly higher than  
observed in the SXDS at $z=3.1$, but slightly below the SXDS at $z=5.7$. 
The difference is up to $\sim 0.5\,\mbox{dex}$, so that the accuracy of 
the model prediction for the LFs should be around $\sim\pm 0.5\,\mbox{dex}$. 
From this catalogue we selected LAEs using similar colour constraints 
to those applied to the observed data,  
see \S \ref{sect:sample} for details of the selection procedure.

\section[]{Sample selection}
\label{sect:sample}
\subsection{The \tnj1338\ field sample}
We used the {\it IA624} image for detection, and the wavelength-weighted 
mean of $B$ and $R$ band images (hereafter {\it BR}) to 
determine the continuum level at a rest-frame wavelength of 1216\AA. 
The {\it BR} image was generated by combining the two images with: 
\[
BR = \frac{(\lambda_{Ly\alpha, \rm obs} - \lambda_B) R +
(\lambda_R - \lambda_{Ly\alpha, \rm obs}) B}
{\lambda_R - \lambda_B}
\]
where $\lambda_B$ and $\lambda_R$ are the central wavelengths of 
$B$ and $R$ bands, respectively, and the $\lambda_{Ly\alpha, \rm obs}$ 
is the observed-frame wavelength of the \lya\ line at $z=4.11$. 
The source detection and photometry were made using the source- 
detection and classification tool, SExtractor \citep{bertin1996}. 
The sources detected here have at least five connected pixels 
above a threshold corresponding to $1.5\sigma$ of the sky noise. 
Using the position of these sources detected in the {\it IA624} image, 
photometry was measured in the other bands with the same aperture, 
after matching the PSF size. We measured photometry for a total of 
205,011 sources, after masking. 

Based on the photometry catalogue from these three images 
($B$, $R$, and {\it IA624}: $i'$ was not used for the selection 
since the data is rather shallow), we selected the LAE candidates 
at $z\simeq 4.1$ by applying the following conditions. 
\begin{eqnarray}
&20 < {\it IA624} < 26.6 \\
&BR - {\it IA624} > 0.3 \\
&B > 27.87, \mbox{ or, } B-R > 2.17\\
&BR - {\it IA624} >  (BR - {\it IA624})_{4\sigma}
\end{eqnarray}
We first applied the magnitude cut (eq.~1) to remove 
bright foreground contaminants (${\it IA624}>20$) and a faint limit 
to remove false detections (${\it IA624}> 5\sigma$). 
Note that the $5\sigma$ threshold here is set according to the SXDS 
data, not the \tnj1338\ data, in order to make a fair comparison with 
the control sample obtained in the SXDS field. The bright threshold, 
20 mag, was determined from visual inspection of the sources. 
This value roughly corresponds to \lya\ luminosities corresponding 
to the brightest high-redshift radio galaxies 
\citep[e.g.,][]{debreuck2001,reuland2003}. 
Eq.~2 selects those sources with \lya\ excess in the intermediate band 
filter, corresponding to an equivalent width of $\ga 200\rm \AA$ 
($\ga 40\rm \AA$ in the rest frame). Then we applied a colour selection 
designed to detect the Lyman-break at $z\sim 4$ (eq.~3). Finally, 
eq.~4 requires that the {\it IA624} excess has a significance level 
of at least $4\sigma$. Note that we used the $BR$ matched-continuum 
to estimate the {\it IA624} excess, although the $B$ band should detect 
almost no flux from $z\sim 4$ sources because the wavelength coverage 
of the $B$ band is mostly below the Lyman limit. However, the distribution 
of the $R-{\it IA624}$ colour is not centred at zero, so that defining 
{\it IA624}-excess sources using just the $R$-band continuum level is 
somewhat unclear. We thus used the $BR$ to measure the continuum levels.

\begin{figure}
\includegraphics[width=84mm]{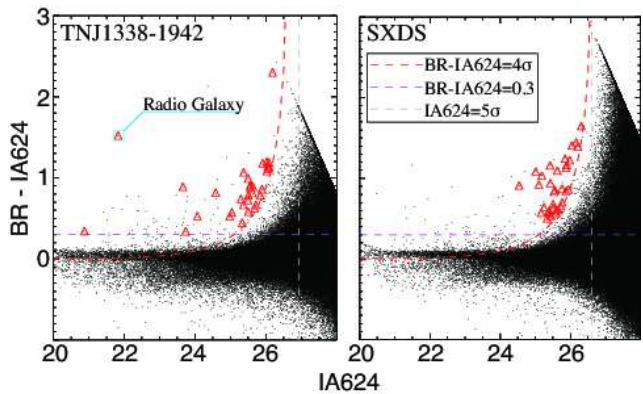}
\caption{Colour-magnitude diagram. 
$BR-{\it IA624}$ colour as a function of  {\it IA624} magnitude for the 
\tnj1338\ field. 
(Left) The black dots show all the sources, and the red triangles show 
$4\sigma$-excess LAE candidates. 
The vertical grey dashed line shows the magnitude cut (${\it IA624}>26.6$), 
the horizontal magenta dashed line indicates the $BR-{\it IA624}>0.3$ 
threshold. The red dashed curves show $4\sigma$ level of the 
{\it IA624} excess. 
(Right) Same as the left panel, but for the SXDS blank field. }
\label{fig:cmd}
\end{figure}

Fig.~\ref{fig:cmd} shows the colour-magnitude diagram used for our 
LAE selection. We need to be aware of the possible contamination 
due to the noise in the {\it IA624} excess measurements: the LAEs 
should lie well above the scatter around $BR-{\it IA614}=0$. 
For some cases, i.e., when the {\it IA624} is relatively shallow, 
false detections may dominate sources selected at the $3\sigma$ level. 
We carefully visual inspected all the sources selected here, 
and we found two source that is obviously affected by bad regions 
of the CCD. These sources are in a region affected by a neighbouring 
bright source. Excluding this, we constructed a sample of 31 LAEs 
in the \tnj1338\ field. 

\begin{figure}
\includegraphics[width=84mm]{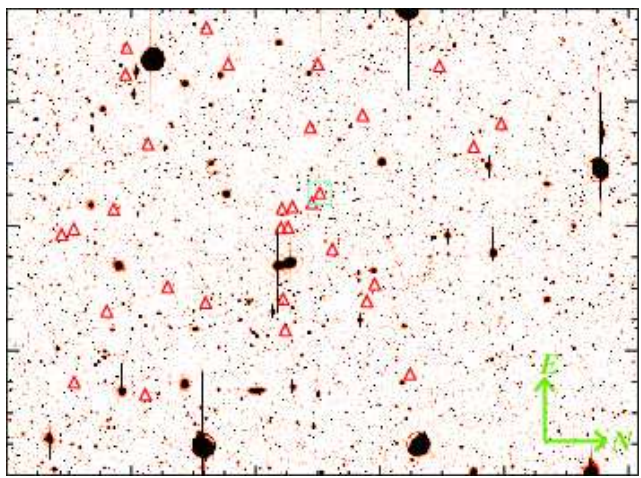}
\caption{The sky distribution of our LAEs in the \tnj1338\ field. 
The symbols for our photometrically-selected LAE sample are 
the same as Fig.\,\ref{fig:cmd}, and the position of \tnj1338\ is 
shown with the cyan  box. The background is the {\it IA624} 
image taken with Suprime-Cam. 
East is up, and north is to the right}
\label{fig:skydist}
\end{figure}

Fig.~\ref{fig:skydist} shows the sky distribution of the LAEs selected 
above, overlaid on the {\it IA624} image. 
The reader should note the apparent ``void''  around the north-western 
part of the field, where relatively few LAEs are detected, and does not 
seem to be an artifact of the bright stars. 

The photometric properties of the sources selected here are 
summarised in Table \ref{tab:list_LAE}. We here checked the 
overlap of our sample with previous studies. We cross-matched 
the coordinates of our sources with LAEs in the sample of 
\citet{venemans2007} and Lyman break galaxies (LBGs) in 
\citet{overzier2008}, both observing the same field. 
We found three sources also included in the previous LAE sample, 
and four with the LBG sample. 
Furthermore, when we apply slightly looser colour constraint, 
$BR-{\it IA624}>3\sigma$ instead of the eq.~4, we have seven 
and eleven overlapped sources with the LAE- and LBG samples, 
respectively. This shows that our colour constraints (eq.~1-4) 
are working at least qualitatively well to select LAEs at $z\sim 4$. 
Since our redshift coverage is $\sim 5$ times wider than that of 
\citet{venemans2002,venemans2007}, most of our sources are 
likely to be located outside the Venemans et al.'s coverage. 
Indeed, only three sources associated with the density peak 
around the radio galaxy (including the radio galaxy itself) 
are overlapped with the Venemans et al.'s sample. 
Our sample thus traces the overdense structure larger than 
$\sim 33 (23\, h^{-1})$ comoving Mpc along the line of sight. 
Altough most of our sources do not overlap with the other samples 
of LAEs and LBGs, their UV ($R-i'$) colours are fairly consistent 
with that expected for $z\sim 4$ galaxies. 
We computed the mean and the standard deviation of the $R-i'$ colours 
of our and Venemans et al.'s samples of LAEs, and obtained the colour 
indices of $-0.14\pm 0.49$ and $0.19\pm 0.38$, respectively. 
Both of them agree with the $R-i'$ colours of our SXDS control sample 
within the errors. 
We then extracted 11 sources with $i'<26.0$ from our sample ($R-i'$ 
colours of $0.08\pm 0.4$), and found that 10 of them satisfy the LBG 
selection criteria by \citet{ouchi2004}. 
The remaining one source has slightly red $R-i'$ 
colour, but fainter than $1\sigma$ level in $B$ band, which is consistent 
with the Lyman break feature. In any case, this source lies in relatively 
less-luminous range of the luminosity function of our sample, and is 
located in average-density region ($\sim 1.3\times$ the average). 
Even if this source is a foreground contaminant, our results thus does 
not significantly change. 
For the sources overlapped with the Overzier et al.'s LBG sample, 
all including the radio galaxy satisfy the colour constraints of 
LBG, showing that our colour selection is working quite well. 
We thus concluded that our sample does not contain significant 
contamination of the foreground sources. 

\begin{table*}
\begin{minipage}{160mm}
  \caption{List of the photometrically-selected LAEs in the \tnj1338\ field}
  \label{tab:list_LAE}
  \begin{tabular}{@{}lccccccccc@{}}
  \hline
ID& $\alpha$ (J2000)& $\delta$ (J2000)& 
{\it IA624}& $BR$& $B^a$& $R^a$& $i'^a$& $\log(L_{\rm Ly\alpha}[\ergs])\, ^b$& Notes\\
\hline
 47564& $13:37:39.77$& $-19:51:57.4$& 26.06& 27.19& 28.62& 26.91& 27.44& 43.21& \\
 54822& $13:37:42.61$& $-19:55:48.4$& 25.55& 26.43& 28.60& 26.16& 26.40& 43.34& \\
 58749& $13:37:44.52$& $-19:37:37.8$& 24.98& 25.49& 28.62& 25.21& 25.36& 43.41& \\
 83374& $13:37:54.64$& $-19:44:22.8$& 25.50& 26.21& 28.62& 25.92& 25.98& 43.31& \\
 93306& $13:37:58.96$& $-19:54:02.5$& 25.33& 25.77& 28.62& 25.49& 24.23& 43.21& \\
 97780& $13:38:00.88$& $-19:48:41.8$& 26.08& 27.19& 28.62& 26.90& 28.31& 43.19& \\
 98384& $13:38:01.22$& $-19:39:58.1$& 25.50& 26.07& 28.62& 25.78& 25.75& 43.23& \\
 99860& $13:38:01.71$& $-19:44:29.6$& 25.37& 26.03& 28.44& 25.75& 25.87& 43.33& \\
106649& $13:38:04.53$& $-19:50:44.9$& 25.29& 26.02& 28.62& 25.74& 26.10& 43.39& \\
107660& $13:38:05.17$& $-19:39:32.6$& 24.06& 24.58& 27.91& 24.30& 24.37& 43.78& \\
127592& $13:38:13.16$& $-19:41:50.0$& 26.19& 28.49& 28.62& 28.23& 26.92& 43.29& \\
135689& $13:38:16.58$& $-19:56:29.3$& 25.72& 26.38& 28.62& 26.09& 26.11& 43.19& \\
138350& $13:38:17.81$& $-19:55:49.6$& 25.04& 25.60& 28.62& 25.32& 25.38& 43.41& \\
139312& $13:38:18.10$& $-19:44:37.0$& 26.09& 27.27& 28.62& 26.95& 27.15& 43.21& \\
139843& $13:38:18.23$& $-19:44:14.1$& 25.69& 26.30& 28.15& 26.06& 26.59& 43.18& \\
149890& $13:38:22.31$& $-19:53:40.2$& 25.50& 26.49& 28.62& 26.23& 26.68& 43.40& \\
150274& $13:38:22.48$& $-19:44:33.7$& 25.51& 26.28& 28.62& 26.02& 25.89& 43.33& {\it d,e}\\
151277& $13:38:22.92$& $-19:43:59.3$& 25.56& 26.46& 28.62& 26.19& 26.63& 43.35& {\it e}\\
153858& $13:38:23.78$& $-19:42:56.8$& 26.10& 27.23& 28.62& 26.91& 26.98& 43.19& {\it e}\\
184585& $13:38:36.63$& $-19:34:11.3$& 25.62& 26.43& 28.13& 26.17& 26.64& 43.29& \\
186397& $13:38:37.39$& $-19:51:48.8$& 25.91& 27.08& 28.62& 26.77& 26.20& 43.28& \\
194782& $13:38:41.12$& $-19:43:01.9$& 23.73& 24.06& 27.31& 23.78& 23.65& 43.75& {\it d}\\
197870& $13:38:41.91$& $-19:32:41.1$& 25.38& 26.44& 28.62& 26.16& 26.61& 43.47& \\
202840& $13:38:43.90$& $-19:40:11.5$& 26.03& 27.11& 28.62& 26.85& 27.00& 43.21& \\
225768& $13:38:53.34$& $-19:53:01.1$& 23.65& 24.53& 27.23& 24.26& 24.27& 44.11& \\
229080& $13:38:55.18$& $-19:36:02.4$& 24.58& 25.39& 28.62& 25.12& 25.33& 43.71& \\
232509& $13:38:55.67$& $-19:42:36.9$& 25.89& 26.74& 28.62& 26.45& 26.25& 43.20& \\
232582& $13:38:55.67$& $-19:47:27.7$& 25.85& 26.62& 28.62& 26.35& 26.64& 43.19& \\
242571& $13:38:59.45$& $-19:52:57.5$& 26.03& 27.21& 28.62& 26.98& 27.05& 43.23& \\
257624& $13:39:04.04$& $-19:48:38.6$& 25.64& 26.52& 28.62& 26.26& 26.83& 43.31& \\
\hline
155683& $13:38:26.07$& $-19:42:30.5$& 21.81& 23.33& 28.62& 23.04& 24.05& 44.97& {\it c,d,e}\\
\hline
\end{tabular}

Notes: ({\it a}) Broadband magnitudes are replaced with the $1\sigma$ values 
when the photometry results are below the $1\sigma$ level; 
({\it b}) \lya\ line luminosities are estimated from the photometry by 
assuming that they are located at $z=4.1$; 
({\it c}) Radio galaxy; 
({\it d}) also included in the LAE sample of \citet{venemans2007}; 
({\it e}) also included in the LBG sample of \citet{overzier2008}. 
\end{minipage}
\end{table*}

The expected contaminants are [O {\sevensize II}]$\lambda 3727$ 
emitters at $z\sim 0.7$, and [O {\sevensize III}]$\lambda\lambda$4959, 5007 
emitters at $z\sim 0.25$. Although our selection corresponds to very large 
EWs, these contaminants cannot be ruled out simply by their EWs, 
as there are certain amount of such strong emitters 
\citep[e.g.][]{vanderwel2011,atek2011}. The LFs of such strong emitters 
are still not studied well, but narrowband surveys can give a rough 
estimate of the number density of such populations. 
\citet{kakazu2007} pointed out that strong [O {\sevensize III}] emitters 
are the most common among such strong smission line galaxies, 
and their [O {\sevensize III}] LF gives the number density of 
the sources with EWs of $\log(\rm EW[\AA]) \ga 2$ to be 
$\sim 1\times 10^{-3}\, \rm Mpc^{-3}$ at $z\sim 0.6$. 
If we assume that this density is valid for $z\sim 0.25$, the expected 
number of [O {\sevensize III}]$\lambda$5007 within our survey volume is 
around unity, when integrating over the whole luminosity range of our sample. 
Even if other emitters ([O {\sevensize III}]$\lambda$4959 and 
[O {\sevensize II}]$\lambda$3727) have similar number density, 
only a few sources would be contained in our sample, which is quite unlikely. 
To test how the contamination affect our results, we reduced the 
number of sources by using $BR-{\it IA624}>5\sigma$ instead of eq.~4. 
This selection excludes one source lying within the high-density 
region around the radio galaxy, giving the peak overdensity well within 
the error. This does not make any significant changes in our results, 
i.e., the density field and the shape of the \lya\ luminosity function. 

\subsection{The control sample in SXDS field}
We constructed a control sample of LAEs, using similar 
imaging data in a blank field (the SXDS-S field, hereafter SXDS). 
These data were taken with the same instrument and filters as used 
in the \tnj1338\ observations as noted above, and the field is not biased 
to any known overdense regions at $z\sim 4$. 
In order to make a fair comparison between the two fields, 
we defined the colour constraints based on the shallower data of the two, 
i.e., $5\sigma$ limiting magnitudes are assumed to be 
${\it IA624}=26.60\,\rm mag$, $R=26.94\,\rm mag$, and $B=26.87\,\rm mag$. 
We also corrected for the offset of $BR-{\it IA624}$ colour distribution 
for the SXDS field, presumably due to an error in the magnitude 
zero point. We then applied a colour term  
$\Delta(BR - {\it IA624}) = 0.15$. This colour term was 
estimated by fitting a Gaussian function to the colour distribution. 
We derived the colour offset necessary to force the centre of the 
colour distribution to zero within the magnitude range 
$20<{\it IA624}<26.60$. 
Except for applying this colour term, we employed exactly 
the same colour constraints to select the LAEs in this field. 
The total number of sources identified is 34. 
The colour-magnitude diagram for this sample is shown in 
Fig.~ \ref{fig:cmd} together with the \tnj1338\ sample. 

\subsection{The mock LAE sample}
\label{sect:mock}
In order to make a meaningful comparison between the observations 
and the simulation, we selected LAEs from the mock LAE catalogue 
at $z=4.17$ described in \S\ref{sect:data}, which is the output redshift 
closest to that of \tnj1338. 
We used the predicted \lya\ fluxes and EWs contained in the catalogue, 
to calculate the colours and magnitudes of the LAEs. 
Note that the catalogue contains $z=4.17$ LAEs, 
while we are going to constrain the observed (apparent) colours of 
$z\approx 4.11\pm 0.12$. However, the difference between these 
two epochs is very small, $\sim 55\,\rm Myr$, and $z=4.17$ is 
well within  the coverage of the {\it IA624} filter. 
We thus calculated the apparent (observed) colours at $z=4.11$. 
This should be a reasonable assumption to compare the simulated 
LAEs with the observations, since the timescale of galaxy evolution 
(e.g., star formation) is much longer than this difference.

We first calculated the absorption of the UV continuum by the 
intervening IGM, following the formalism of \citet{madau1995}. 
Here the UV continuum was assumed to be flat in terms of 
flux density per unit frequency bin, $f_\nu = \mbox{\rm Const}$. 
The flux density was obtained by dividing the \lya\ flux 
by EW (observed), $f_\lambda = F({\rm Ly\alpha}) / W_{obs}$. 
Then the IGM absorption was calculated by assuming the 
redshift of $z=4.11$ to obtain the continuum component of 
the spectrum expected at the redshift of the radio galaxy. 
The UV continuum contribution to the photometry 
was calculated by convolving the flat-continuum spectrum 
with the filter response curves. 
The IGM absorption on the \lya\ line is included in the 
escape fraction, $f_{esc}$ used in generating the mock catalogue of LAEs 
\citep[see][]{orsi2008}. 
Then sufficiently narrow line profiles (i.e., narrower than the 
{\it IA624} passband) was convolved with the filter 
response curves, and the line contribution to the photometry 
was obtained. 

Adding the contributions from both the \lya\ line and the 
UV continuum, we calculated the $R$- and {\it IA624} band magnitudes 
to apply the colour constraints. Since our calculation gives 
exactly the same continuum colour for all sources, we 
did not apply the constraints on the $B-R$ colour. 
This does not affect our results because the mock LAE 
catalogue does not contain any foreground contamination. 
Furthermore, the $B$-band magnitude cannot be predicted 
from our calculation with sufficient reliability, 
because we are assuming a very simple IGM absorption model. 
Due to the Lyman-break feature and the Gunn-Peterson trough, 
the high-$z$ sources are expected to be fainter in $B$ than in the $R$ band. 
We indeed required in eq.~3 that $B-R>2.17$, which leads to be 
$\ga 0.15$ fainter $BR$ magnitudes than the $R$. 
In order to put the same colour- and magnitude constraints as for the 
observed samples, we then put the offset of $+0.15$ mag on the $R$-band 
magnitudes when computing the {\it IA624} excess, 
instead of using $BR$ magnitudes. This gives the following 
criteria for selecting the LAEs: 
\begin{eqnarray}
&20 < {\it IA624} < 26.6 \\
&R - {\it IA624} > 0.15 \\
&R - {\it IA624} > (BR - {\it IA624})_{4\sigma}.
\end{eqnarray}
For the significance level of the {\it IA624} excess, 
$(BR-{\it IA624})_{4\sigma}$, the same value as in  
eq.~4 was used to match the observed sample. 
For the comparison with the observed samples, 
we applied the magnitude cut of ${\it IA624}<26.6$ 
for eq.~5, and the observed {\it IA624}-excess threshold of 
$(BR-{\it IA624})_{4\sigma}$. 
In total we select 59,639 model LAE sources  
in a comoving volume corresponding to $(500\, h^{-1}\rm Mpc)^3$. 

\section[]{Results and discussion}
\label{sect:results}
\subsection{Density field}
\label{sect:d_comp}
\begin{figure*}
\includegraphics[width=160mm]{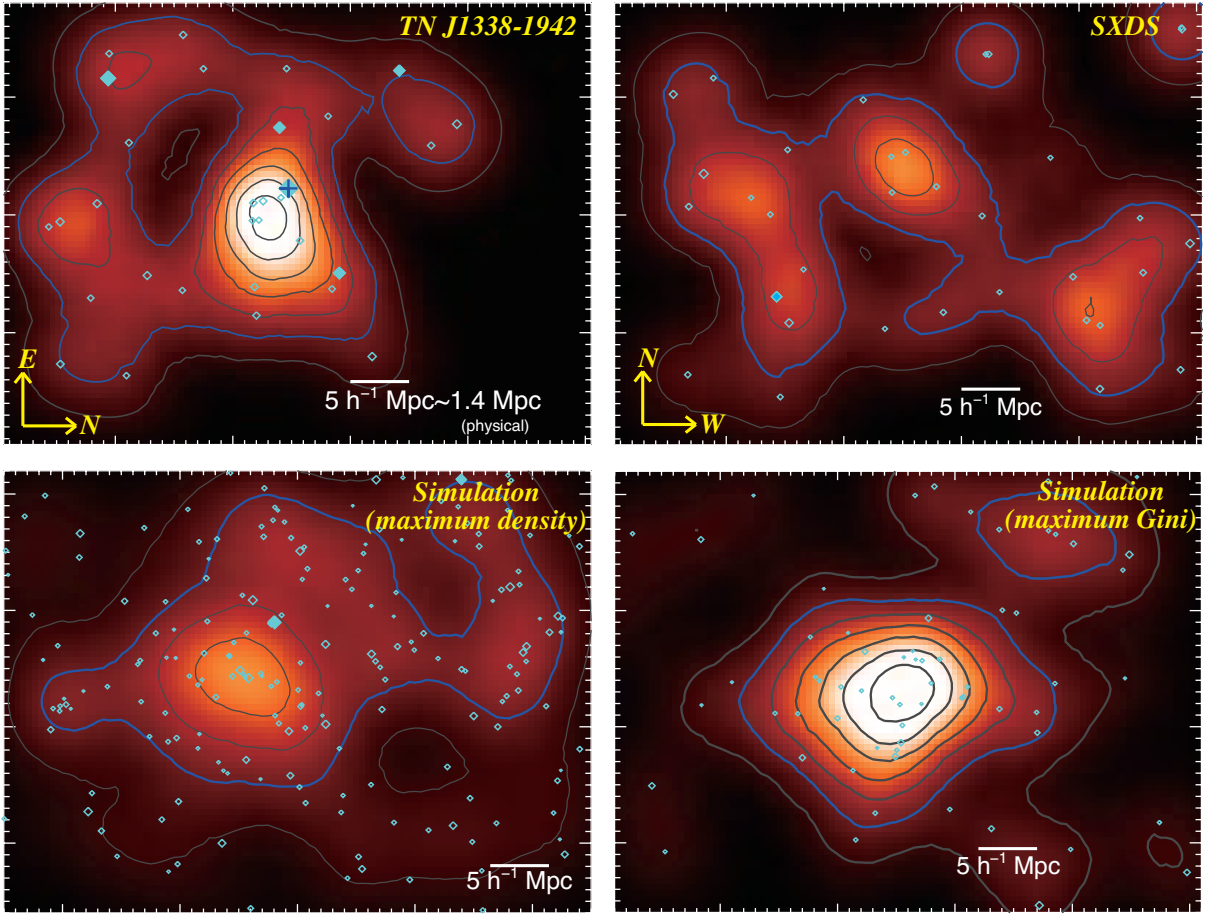}
\caption{A comparison of the density field of LAEs. 
The top panels are the observed density map of the \tnj1338\ field 
(left), and the SXDS field (right). 
The position of the radio galaxy \tnj1338\ is shown with the blue 
cross mark near the density peak. 
The bottom panels are simulated density maps generated from the 
mock LAE catalogue using the same selection as those applied to 
the observed data, showing the same FoV centred on density peaks 
which are higher than that measured at the position of the radio galaxy: 
the field around the maximum density peak (left) and 
the field with the highest Gini coefficient (right). 
Each panel shows the $50 \times 38\, h^{-2}\rm Mpc^2$ (comoving) field. 
The contour  interval is  $0.5\times$ the average density over 
the same FoV, with a lowest contour of 0.5 ($\delta_{\rm LAE}= -0.5, 0.0, 
0.5, 1.0, \dots$). 
The thick blue contour traces the average density over the field 
($\delta_{\rm LAE}=0.0$). 
The cyan diamonds mark the positions of LAEs, and the sizes of these 
symbols represent the \lya\ luminosities. 
The sources with $\log(L_{\rm Ly\alpha}[\ergs])\ga 43.6$ are marked 
with filled symbols. 
}
\label{fig:dmap_comp}
\end{figure*}

\subsubsection{The high density region around the radio galaxy}   
\label{sect:hdr}
We quantified the difference between the \tnj1338\ and SXDS fields 
by constructing their LAE surface density maps. 
We smoothed the spatial distribution of the LAEs with 
a Gaussian kernel with a radius (half width at half maximum, HWHM) 
of $4\, h^{-1}\rm Mpc$ (projected comoving distance at $z=4.1$, 
corresponding to the physical distance of 1.1 Mpc), 
and counted the number of LAEs within the same radius 
at each grid point, with a grid spacing of $0.5\, h^{-1}\rm Mpc$. 
Then the surface density was obtained by dividing 
the number by the area of the Gaussian aperture 
with the HWHM of $4\, h^{-1}\rm Mpc$. In deriving the average 
surface density, the areas within $5\, h^{-1}\rm Mpc$ of the 
edge of the FoV were flagged to avoid underestimating 
the surface density near the field edges. 
The kernel size was chosen based on the typical separation 
between the sources. We measured the distance to the nearest 
source for each LAE in our sample, and chose the smoothing kernel 
size to include $\sim 68$ percent of the whole sample. 
We confirmed that different sizes of the smoothing kernel 
also give a similar density distribution. 
Note that the radio galaxy itself was excluded in creating the 
density map, since we are aiming to investigate the surroundings 
of the radio galaxy.

The density maps we obtained are shown in the Fig.~\ref{fig:dmap_comp}. 
We can clearly see that the region around the radio galaxy 
is strongly overdense, and equivalently strong density peaks are not  
seen in the SXDS field. The density peak in the \tnj1338\ field 
is $\approx 4$ times the density averaged over the field, 
while the maximum peak in the SXDS field is only $1+\delta_{\rm LAE}=2.36$. 
We then estimated the uncertainty of the density with this 
smoothing scale by computing the standard deviation of the 
density map of the SXDS field, determining the $1\sigma$ dispersion 
in $\delta_{\rm LAE}$ as 0.47. 
The radio galaxy \tnj1338\ is located in the densest region in 
the field, with $1+\delta_{\rm LAE}=2.91$ at its position. 
The peak density is $1+\delta_{\rm LAE}=3.80$, located 
$2.8\, h^{-1}\rm Mpc$ offset from the position of the radio galaxy. 
This overdensity can be traced on $\sim 10$--$20\, h^{-1}\rm Mpc$ scales 
around the radio galaxy ($\sim 3$--$6\,\rm Mpc$ in physical). 
Outside this region, in contrast, the density appears to drop quite 
rapidly: on the northwestern side of the field there is a strongly 
underdense region just $\sim 20\, h^{-1}$ Mpc from the density peak. 
Such a strong variation in the density of LAEs is not seen in the SXDS 
field. Although the overdensity of the \tnj1338\ field had previously been  
suggested by \citet{venemans2002}, their FoV was much smaller than ours, 
and they were forced to estimate the overdensity (a factor of $\sim 3$--5) 
through comparison to a separate field survey at a similar redshift 
\citep{rhoads2000}. With our wide-field imaging of this field, we have 
not only confirmed the overdensity by comparing with an identically-observed 
control field, but have also determined the spatial scale and structure 
of the overdense region. 

\begin{figure*}
\includegraphics[width=160mm]{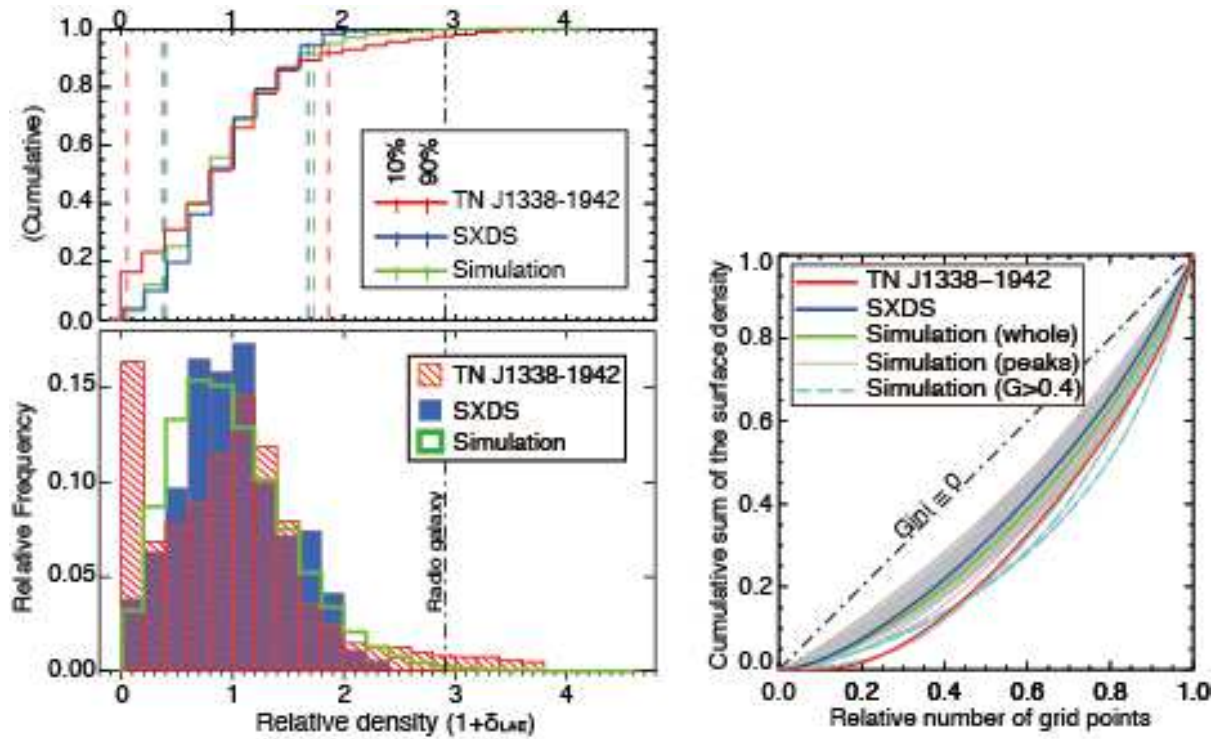}
\caption{The statistics for the LAE surface density distribution. 
(Left) A comparison of the LAE surface density distributions for 
the two fields. The red hatched and the blue filled histograms show the 
density distribution for the \tnj1338\ field and the SXDS control 
field, respectively. Overplotted with the thick green line 
is the histogram for the simulated density map obtained from the 
mock LAE sample. The vertical black dot-dashed line shows the 
density calculated at the position of the radio galaxy. 
The cumulative distributions are shown on the top panel. 
The red, blue and green lines denote the \tnj1338 , the SXDS and 
the simulated map, respectively. 
The 10-th and 90-th percentiles are shown with the vertical dashed lines 
with the corresponding colours. We can see that the density distribution 
in the \tnj1338\ field is much wider than that in the SXDS. 
The simulated density distribution is closer to the SXDS than 
to the \tnj1338\ field. 
(Right) Lorentz curves for the density distribution. 
The normalised cumulative distribution is plotted against the normalised 
number of the grid points. The red, blue and green curves show 
the \tnj1338, the SXDS and the simulated map, respectively. 
Again the simulated distribution is closer to that of the SXDS 
field rather than the \tnj1338\ field. 
The grey curves denote the 94 fields centred at the high-density peaks 
found in \S\ref{sect:comp_whole}, of which 21 have $G$'s higher than 
that of the whole simulated map. 
The three fields with the largest $G$'s among the 94 ($G\ge 0.4$) are 
shown with the cyan dashed curves. 
The \tnj1338\ field has much higher $G$ than most of the high-density 
fields found in the simulation: only two out of 94 has higher $G$'s 
compared with the \tnj1338\ field. 
}
\label{fig:hist_comp}
\end{figure*}

The difference between the \tnj1338\ field and SXDS control field 
becomes clearer when comparing their density histograms, as shown 
in Fig.~\ref{fig:hist_comp}. 
In the SXDS field the distribution is concentrated near the average value, 
while the \tnj1338\ field shows a much broader distribution extending 
toward both high- and low densities. 
In order to quantify the difference between the distributions in these 
two fields, we first compared the 10- and 90 percentiles for the 
density maps of the two fields. The upper panel of 
Fig.~\ref{fig:hist_comp} shows the cumulative density 
distributions for the two fields, together with the percentiles. 
The statistics of the two fields are also summarised in 
Table~\ref{tab:stat_obs}.
The \tnj1338\ field has a wider density distribution compared with 
the SXDS field, especially at the lower-density end. 
The 10 percentile for the \tnj1338\ field is $1+\delta_{\rm LAE}=0.05$, 
which is eight times lower than that of the SXDS field. 
The ratios of the two (10 and 90) percentiles, i.e., the dynamic range of the 
density, for the \tnj1338\ and the SXDS fields are 37 and 4.2, respectively. 
The \tnj1338\ field has thus nearly an order of magnitude higher range 
in galaxy density than that seen in the SXDS field. 
The difference becomes even more obvious when we look into the highest- 
and the lowest-density bins: e.g., $1+\delta_{\rm LAE}<0.01$ corresponds 
to the lowest 9.5 percent of density cells in the \tnj1338\ field, while 
in the SXDS field only $\sim 0.2$ percent of the total grid points have 
such a low density. 

A useful statistic to quantify the range of the density distribution is 
the Gini coefficient of the density field. The Gini coefficient, $G$, 
measures how uniformly LAEs are distributed within the field, and takes 
a value of $0<G<1$. The $G$ value corresponds to the area surrounded by 
the two Lorentz curves corresponds to the given distribution and $G=0$ 
(shown in Fig.~\ref{fig:hist_comp}). 
If all the LAEs are concentrated within one grid point, then $G$ is unity. 
If the LAEs are distributed uniformly over the field, then $G$ is zero. 
The Lorentz plot shown in Fig.~\ref{fig:hist_comp} 
clearly shows that the \tnj1338\ field has much higher $G$ than the 
SXDS blank field, and the blank field agrees well with the simulation. 
The Gini coefficients of the density maps were calculated to be 
0.402 (0.268) for the \tnj1338\ (SXDS) field. This difference 
shows that the LAEs are more concentrated into high-density 
regions (especially into the high-density peak around the radio galaxy) 
in the \tnj1338\ field, compared with the SXDS field. 
Together these statistics suggest that the galaxy density in the \tnj1338\ 
field traced with LAEs is highly concentrated within the high-density 
region near the radio galaxy, but that there are also unusually low-density 
regions in the field where the number of LAEs is highly suppressed. 
This is quite different from the density field in the SXDS field.

\begin{table}
\centering
\begin{minipage}{85mm}
  \caption{Statistics of the observed density field}
  \label{tab:stat_obs}
  \begin{tabular}{@{}ccccccc@{}}
  \hline
 Field & Mean $^a$& Peak $^b$& \multicolumn{3}{c}{Percentiles $^c$}& $G$ $^d$\\
 & $[h^{2}\rm Mpc^{-2}]$ &  & 10 & 50 & 90 & \\
\hline
 \tnj1338 & 0.0201 & 3.80 (3.73) & 0.05 & 0.98 & 1.86 & 0.402 \\
 SXDS     & 0.0204 & 2.36 & 0.40 & 0.97 & 1.68 & 0.268 \\
\hline
\end{tabular}

$^a$Mean surface density over the field in comoving scale.\\
$^b$Peak normalised surface density ($1+\delta_{\rm LAE}$). 
The value normalised with the average of SXDS field is shown in 
the parenthesis.\\
$^c$In units of the normalised density $(1+\delta_{\rm LAE})$.\\
$^d$Gini coefficient of the density distribution.\\
\end{minipage}
\end{table}

\subsubsection{Density contrast within the \tnj1338\ field}
\label{sect:d_contrast}
Although the \tnj1338\ field has a very large dynamic range in  
LAE density, the average density of this field over the FoV of 
Suprime-Cam is still almost the same as that of the SXDS field. 
The average surface density of LAEs for these two fields are 
$0.0201\, h^{2}\rm Mpc^{-2}$ and $0.0204\, h^{2}\rm Mpc^{-2}$ 
for the \tnj1338\ and the SXDS field, respectively. 
This suggests that the density traced by LAEs is 
almost uniform (within $\sim 2$ percent) at $\sim 50\, h^{-1}\rm Mpc$ scales, 
when the density field is averaged over $\sim 112\, h^{-1}\rm\, Mpc$ 
along the line of sight (both in comoving scale). 
Even with this smoothing, there is a significant overdensity 
around the radio galaxy, and a similarly significant underdense region 
just adjacent to it. This ``void'' region next to the 
high-density region corresponds to the peak of the density 
distribution near zero in the histogram (Fig.~\ref{fig:hist_comp}). 
The shape of this distribution is quite different from that for 
the SXDS field. 

Due to the limited number of sources, $\sim 30$ in each field, 
it is not clear whether the underdense region is truly a void. 
However, the detectable (i.e., bright and large-EW) LAEs 
were not found in this region, especially in the north-western quarter of 
the field. This appears to be a real effect since there are no bright 
stars in this region that would significantly affect the detection and 
photometry, as seen in the Fig.~\ref{fig:skydist}. Such a large 
underdense region is not seen in the SXDS field, again showing that 
this radio galaxy field has an unusually high density contrast. 
The real density contrast in this radio galaxy field is possibly 
higher than estimated, since we are smoothing the spatial distribution 
along the line of sight as a result of the relatively wide redshift 
coverage of the {\it IA624} filter. Although the apparent contrast 
would be enhanced due to the redshift-space distortion if the high-density 
region is actively accreting the material, such an effect is thought to be 
small because of the wide redshift coverage. 

Such a large density contrast within a single field suggests that 
the high-density region around the radio galaxy is attracting material 
from well within the scale of the Suprime-Cam FoV ($38\times 50\, h^{-2}$ 
comoving Mpc$^2$ on the sky, when the edges are flagged out), because 
the average density is almost the same as the blank field. If LAEs are 
tracing the matter distribution, then the surrounding dark haloes within 
this scale should have been merging into this overdensity. Since we do 
not have spectroscopic data for all the sources, we cannot draw any clear 
conclusions on the true spatial structure of the overdensity based 
solely on our current observations: it is not clear whether the 
density peak represents a single extremely overdense halo, or a less  
overdense filamentary structure elongated toward the line of sight 
(the latter case includes the case that two or more clumps are aligned 
along the line of sight). Nevertheless, the density contrast must 
have grown to a level high enough to form such a high density 
peak and a large void region. Even for the latter case, the high-density 
filament must have accreted material from its surroundings and grown  
to a sufficient length to evacuate the void region. Hence in both cases, 
the spatial extent of the high-density region should be fairly compact 
(or the filament should be narrow), as the \tnj1338\ field shows a large 
void fraction and the high-density region extends only up to 
$\sim 20\, h^{-1}$ comoving Mpc ($\sim 6$ Mpc in physical units) scales. 

This requires that matter is concentrated into a high-density region of 
$\sim 10$--$20\, h^{-1}$ comoving Mpc size ($\sim 3$--6 physical Mpc), 
well before the observed epoch of $z\sim 4.1$. Such a concentration 
should affect the star-formation and AGN activity of galaxies within 
the overdensity, in the form of, e.g., frequent galaxy mergers. 
This may therefore lead to an excess of massive galaxies harbouring 
active star-formation and/or AGN activity. 
The corresponding high matter concentration will also lead to a higher 
rate of gas accretion from the surrounding environment, again resulting 
in the enhancement of star-formation and AGN activity. 

\subsection{Luminosity function}
\label{sect:lf}
\subsubsection{Comparison with the blank field}
\begin{figure}
\includegraphics[width=84mm]{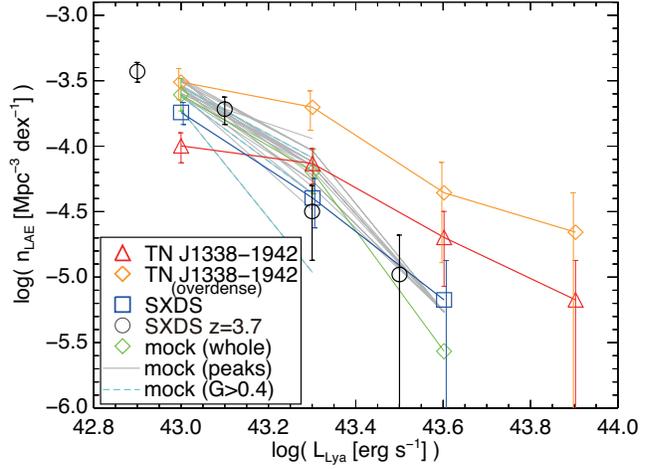}
\caption{
The \lya\ luminosity function (LF) of the observed samples of LAEs. 
The red triangles show the LF of the total LAE sample in the 
\tnj1338\ field. The subsample lying within the overdense regions 
(higher than the average of the field) is shown with the orange 
diamonds. The blue squares denotes the SXDS control sample. 
The black circles shows the LF of $z\simeq 3.7$ LAEs in 
the SXDS field, taken with a narrowband filter \citep{ouchi2008}. 
The green diamonds show the LF of the mock LAE catalogue 
by \citet{orsi2008}, and the grey lines show the LF derived 
from the subsample of the mock LAEs residing within the Suprime-Cam-sized 
fields centred at the positions of 21 high-density peaks with high $G$'s. 
The LFs of the three highest-$G$ peaks among these 21 are marked with 
the cyan dashed lines. 
The radio galaxy \tnj1338\ is omitted to show the luminosity range 
of the normal LAEs. 
Note that the data points are slightly shifted to improve the visibility. 
We can see that the number of bright LAEs is enhanced in the 
radio-galaxy field, and the fraction of the faint LAEs is 
reduced. The LF within the overdense regions has almost the 
same shape as that within the whole field, but the absolute number density 
agrees with that in the SXDS at a luminosity of 
$\log(L_{\rm Ly\alpha}[\ergs])\sim 43.0$. 
}
\label{fig:lf_obs}
\end{figure}

As described above, we have found that the \tnj1338\ field has a
high density region around the radio galaxy, and a strong density 
contrast between this and the large void region just adjacent to the 
peak. To investigate the variation in the LAE LF Fig.~\ref{fig:lf_obs} 
compares the \lya\ LFs of the LAEs in the \tnj1338\ and the SXDS fields. 
We can see that the faint-end slope of the LF in the \tnj1338\ 
field is significantly shallower than the SXDS, below a \lya\ luminosity 
of $\log (L_{\rm Ly\alpha}[\ergs]) \sim 43.3$. 
On the other hand, the LF in the SXDS field increases nearly monotonically 
down to our completeness limit, $\log L_{\rm Ly\alpha}\sim 42.8$, 
as the $L^*$ is still slightly below the faintest data point 
\citep[][hereafter O08]{ouchi2008}. 
In the \tnj1338\ field, a similar trend can be seen in the LF for 
LAEs within the overdense region. 
We plotted the LF of the subsample of LAEs lying within regions 
with densities higher than the average of the whole Suprime-Cam field 
($\delta_{\rm LAE}\ge 0$). 
The faint end of this LF agrees with that of the SXDS field, 
but has higher values than in the SXDS by up to an order of magnitude. 
Note that the radio galaxy itself ($\log L_{\rm Ly\alpha}\approx 45.0$) 
is excluded in the LFs plotted here. 
The LAE population in the \tnj1338\ field is thus thought to be 
highly biased to bright sources, and the fraction of faint LAEs 
is reduced in this field. The shape of the bright end may suggest 
that our LAE sample consists of two different components, 
e.g., star-forming galaxies dominating the faint end, 
and AGN hosts dominating the bright end. 
This should be real even if the bright end contains foreground 
contamination. As mentioned in \S\ref{sect:sample}, the luminosity 
function of foreground sources suggests that the expected number 
of [O {\sevensize II}] emitter is around unity, even when integrated 
down to our completeness limit. The number of LAEs with 
$\log(L_{\rm Ly\alpha})\ga 43.6$ is four in the \tnj1338\ field 
(excluding the radio galaxy), while that in the SXDS field is 
only one. It is unlikely that more than one such bright foreground 
sources are included in our sample, and thus the enhancement of 
the bright LAEs is thought to be real.

We have also compared the SXDS control sample with a narrowband-selected 
LAE sample at a similar redshift in the same field, to check 
the reliability of our control sample. 
The LF of the SXDS control sample agrees quite well with 
the LF of the narrowband-selected LAE sample at $z\simeq 3.7$ (O08), 
down to the completeness limit ($\log(L_{\rm Ly\alpha})\sim 42.8$). 
Since the bandpass of the {\it IA624} filter is several times 
larger than the {\it NB570} narrowband filter, the survey volume 
is comparable to that of O08, even with only one FoV of Suprime-Cam. 
This reduces the field-to-field variance, as the density fluctuations 
are smoothed out along the line of sight. 
Thus we confirmed that our SXDS sample is a valid control sample in terms 
of \lya\ LF, representing the typical number density of LAEs at $z\sim 4$. 
The LF in the \tnj1338\ field, on the other hand, 
lies beyond the variance expected from the results of O08. 
At the brightest end of the LF, $\log L_{\rm Ly\alpha}\ga 43.9$, 
we did not find any sources in the SXDS field, while we still found 
sources in the \tnj1338\ field even when we exclude the radio galaxy itself. 
Such bright sources are almost never seen in blank-field 
surveys for LAEs at similar redshifts \citep{dawson2007,ouchi2008}. 

These results reinforce the idea that the \tnj1338\ field is unusual, 
not only in terms of the overdensity of LAEs, but also the \lya\ LF. 
Since our {\it IA624} data is deeper in the \tnj1338\ field than in 
the SXDS field, this should not be due to the difference in the 
completeness. We mentioned in \S \ref{sect:d_contrast} that the 
density contrast in this field implies that the high-density 
region is accreting material from well within $\sim 50\, h^{-1}\rm Mpc$ 
scale. At this scale, such accumulation of the material is likely to be 
affecting the star-formation and/or AGN activity in galaxies and 
hence galaxy formation and evolution, leading to the enhancement 
of the bright end of the \lya\ LF. The bright end of the \lya\ LF 
is thought to be dominated by AGN hosts and/or actively star-forming 
galaxies, so that the formation of such ``active'' galaxies is likely 
to be enhanced in the \tnj1338\ field.

\subsubsection{Implications for galaxy formation}

The high-density region around the radio galaxy is so unusual that 
it faces the large void region. The material originally in the void 
region must have moved into the surroundings by the epoch of 
$z\simeq 4.1$, and among the ``surroundings'', the high-density region 
around the radio galaxy is the most prominent peak. This suggests 
that some large fraction of the material originally in the void 
region may have travelled to the vicinity of the radio galaxy. 
Then, how well can this scenario account for such a high density 
contrast as we found? We made a rough estimate to test this scenario. 
The separation between the density peak and the void region 
is $15-20\,h^{-1}\rm Mpc$ in comoving scale, which corresponds 
to $\sim 4-6\,\rm Mpc$ in physical scale at $z=4.1$. 
If the material travelled this distance within a Hubble time at 
this redshift, the velocity is required to be 
$\sim 4-6\times 10^2\rm km\, s^{-1}$. 
This is altough a very naive estimate, as is just the velocity 
required to travel linearly from the centre of the void to the 
density peak. In terms of the void evolution, the required 
velocity is slightly smaller: the apparent radius of the 
completely empty void is $r\sim 2-3\,\rm Mpc$, leading to the 
peculiar velocity at the edge of the void is around 
$\frac{1}{3}H(t)r \sim 2-3\times 10^{2}\rm km\, s^{-1}$. 
These values are well below the typical velocity dispersion 
of the present-day (virialised) galaxy clusters. 
Although the required peculiar velocity may be larger than this 
because the distance assumed here is just the projected distance, 
the high-density region can be to some extent responsible for 
accreting the material initially in the void region. 

It is however quite unlikely that all of the material initially in the 
large void had been simply accreted onto the high-density region by 
the epoch of $z\simeq 4.1$. The material should in principle flow out 
in all directions from the centre, so that the single overdense region 
cannot be fully responsible for the formation of the void. 
The simplest interpretation for the observed (apparent) density 
contrast is that the initial density fluctuation is quite large, 
and the contrast had existed since an epoch well before $z\simeq 4.1$. 
Another interpretation is the enhancement of star-formation 
activity within overdense regions \citep[e.g.][]{steidel2005,koyama2013}, 
leading to the enhancement of actively star-forming galaxies possibly 
dominating the bright end of the LF. This may result in a drastic 
enhancement of the galaxy bias for bright sources. 
The excess of extreme starbursts within protoclusters found in 
some observational studies \citep[e.g.][]{blain2004,capak2011,ivison2013}
is consistent with this idea. 
Because of the lack of spectroscopic data, we cannot discriminate 
between star-formation and AGN activities. It is thus also possible 
that AGNs are dominating the bright end of the LF, which is highly 
enhanced in the radio galaxy field. Such an enhancement of AGN activity 
in overdense environments has been found observationally in some cases 
\citep[e.g.][]{pentericci2002,croft2005,lehmer2009,digby-north2010}. 
Since our samples are relatively biased to bright LAEs, 
such effects will enhance the apparent density contrast.

The difference between the LFs in the two fields depends strongly on 
the luminosity. When we examine the LF for the overdense regions 
in the \tnj1338\ field, the difference for the luminosity range  
$\log L_{\rm Ly\alpha}\ga 43.3$ (corresponding to the brightest 
two data points for the SXDS field) is about an order of magnitude. 
This luminosity range contains 15 sources for the \tnj1338\ field 
(all in the overdense regions), while only 7 lie within this range 
in the SXDS field. 
On the other hand, both LFs agree with each other within their 
error bars (factor of $\sim 1.6$) at $\log L_{\rm Ly\alpha}\sim 43.0$. 
The difference in the faintest data point roughly corresponds 
to the difference in the total number density for the two fields 
(the whole SXDS field and the overdense region of the \tnj1338\ field). 
We estimated that the mean density within the overdense regions 
shown here is $\sim 1.6$ times the average of the SXDS, 
corresponding to $\delta_{\rm LAE}\sim 0.6$. 
Since the LF for the overdense regions in \tnj1338\ exceeds the 
LF in the SXDS field by an order of magnitude on the bright end, 
the overdensity of the bright ($\log L_{\rm Ly\alpha}\sim 43.6$) 
LAEs is $\delta_{\rm LAE}\sim 5$ or so. 
This means that, if we assume the galaxy bias for LAEs at 
$\log L_{\rm Ly\alpha}\sim 43$ to be $b\sim 2.4-4$ 
\citep[e.g.][]{ouchi2005,ouchi2010,kovac2007,chiang2013}, 
and the faint LAEs are correctly tracing the matter density of 
this field, the bias for brighter LAEs must be $b\sim 20-40$. 

Such a strong luminosity dependence of the bias seems to be 
rather different from the prediction of the model of \citet{orsi2008}, 
which predicted only a modest dependence of bias on LAE luminosity. 
This may suggest that the bright end of the LF is dominated by 
populations other than the normal star-forming galaxies included in 
that model, such as hosts of AGN or AGN-induced star-formation 
activities. However, it is still not clear if this result is truly 
against the prediction. The bias measurement of \citet{orsi2008} 
is based on clustering on large scale, so that the results are 
different from ours. Furthermore, their Fig.~6 shows large uncertainty 
on the bias at the very bright end of the model LAEs, 
$\log L_{\rm Ly\alpha}\sim 43$, and it is still possible that 
the bias of such bright LAEs depends strongly on the luminosity, 
even for the model LAEs. \citet{orsi2008} did not present predictions 
of the clustering bias for LAEs brighter than $\log L_{\rm Ly\alpha}\sim 43.0$, 
because their simulation did not contain enough such LAEs for an 
accurate measurement on large scales. 

We here showed only the simplest estimate of the galaxy bias, 
since we are heavily affected by small number statistics due to having 
only $\sim 30$ LAEs in each field. We cannot probe the scale-dependence 
of the galaxy bias either: the bias estimated above is on a comoving 
scale $\sim 20-30\, h^{-1}\rm Mpc$, which corresponds to the area inside 
the contour of  average density in Fig.~\ref{fig:dmap_comp}. 
Note again that this is smoothed over $\sim 112\, h^{-1}$ comoving Mpc 
along the line of sight. We can then qualitatively say that there 
apparently exists a large density contrast within $\sim 50\, h^{-1}$ 
comoving Mpc scale around the radio galaxy \tnj1338, apparently enhanced 
to some extent by highly biased nature of the bright LAEs including 
AGN hosts.

\subsection{Comparison with the mock catalogue}
\label{sect:comp_mock}
\subsubsection{Density field of the whole simulated map}
\label{sect:comp_whole}
We compared the density field shown in \S\ref{sect:d_comp} 
with the mock LAE sample, and evaluated the density distributions 
of the \tnj1338\ field, as well as the control field. 
The simulated density maps were obtained by applying the same 
analyses to the mock LAE catalogue described in \S\ref{sect:mock}. 
We used the same smoothing radius, $4\, h^{-1}$ comoving Mpc, in this process. 
The simulation covers a comoving volume of $(500\,h^{-1}\rm Mpc)^3$. 
Our {\it IA624} filter covers a radial comoving distance of 
$\approx 112\, h^{-1}\rm Mpc$, so that we can obtain four 
independent slices perpendicular to the line of sight to create 
mock sky distributions of LAEs. 
After selecting the LAEs with similar colour constraints 
to the observed samples, we extracted the LAEs within a slice 
of thickness (distance along $z$-axis of the simulation box) 
of $112.4\, h^{-1}\rm Mpc$, and projected them onto the $xy$-plane. 
In this process we took account of the effect of redshift-space 
distortions. The redshift-space coordinates were computed by taking 
the $z$ axis as the line-of-sight direction, adopting the distant 
observer approximation. We then smoothed the spatial distribution of 
LAEs projected on the $xy$-plane, and counted the LAEs within a 
smoothing radius centred at each grid point. 
The simulated LAE density map covers $\sim 100$ times larger area 
for each slice than a single pointing of the Suprime-Cam, and thus 
$\sim 400$ times larger volume for the four slices together. 

We first used this whole simulated map to compare the density 
distribution with the observational data. 
Fig.~\ref{fig:hist_comp} shows the density histogram of the whole 
simulated map, overlaid on the observed density histograms. 
We see that the density distribution of the simulated map is at least 
qualitatively consistent with that of the SXDS field. The distribution 
is close to Gaussian with a slightly longer tail toward the high density, 
peaked around the mean density. The peak amplitude and the width of 
the peak both look close to that of the control sample. At higher 
density, on the other hand, the simulated map has a longer, more 
extended tail than that of the control sample. This looks still 
consistent with the control sample, because the field coverage 
of the observational data is much smaller than the simulation. 
However, the high density tail seen in the \tnj1338\ field is larger 
than that in the whole simulated map.

Statistical tests also support this difference between the SXDS 
and \tnj1338\ fields, in terms of the comparison with the simulated map. 
The 10- and 90 percentiles of the density in the simulated map are 
0.37 and 1.73 respectively, and the ratio between the two percentiles 
is 4.7, which is closer to that in the SXDS (4.2) than in the \tnj1338\ 
field (37.3): see Fig.~\ref{fig:hist_comp} left and Table~\ref{tab:stat_obs}. 
We also calculated the Gini coefficient of the simulated 
density map, $G = 0.302$, and compared with the observed 
distributions in the two fields (Fig.~\ref{fig:hist_comp} right). 
The dynamic range of the LAE density for the simulated map is thus 
in between those of the two observed fields, and closer to the SXDS 
field than the \tnj1338\ field. We then assume that the SXDS field 
has a typical density distribution of LAEs at $z\sim 4$, and we can 
evaluate how rare the high-density region seen in the \tnj1338\ field is. 
The density measured at the position of the radio galaxy is 
$1+\delta_{\rm LAE}=2.91$. Such high density regions are quite rare 
in the simulated map, corresponding to the densest $0.07$ percentile 
of the whole volume, while the same overdensity corresponds to the 
$2.9$ percentile of our survey volume in the \tnj1338\ field. 

\begin{figure}
\includegraphics[width=84mm]{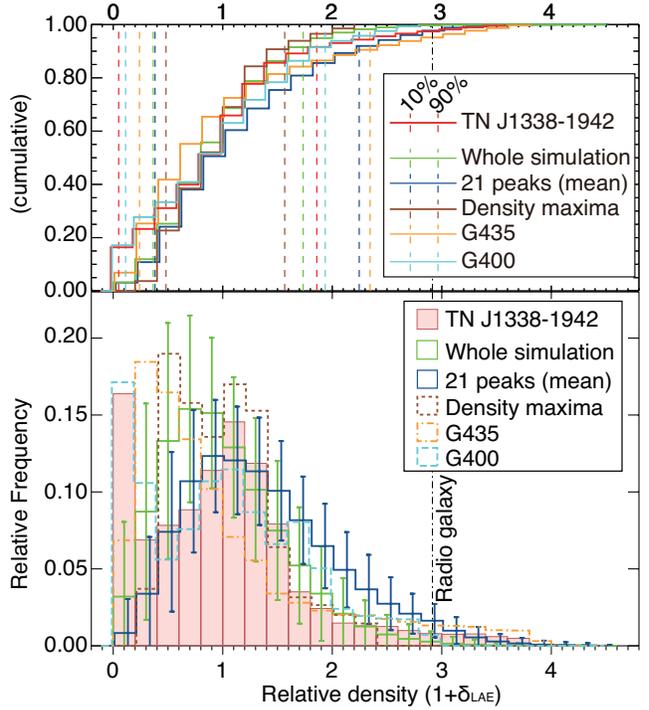}
\caption{
Comparison of the density distributions between the \tnj1338\ 
field and the simulated map with various conditions for the simulation. 
The filled red histogram shows the observed density distribution of the 
\tnj1338\ field (the same as in Fig.~\ref{fig:hist_comp} left). 
The green outlined histogram is the density distribution for the 
whole simulation volume. The errors were estimated by computing the 
density distributions for 400 different Suprime-Cam-sized fields 
centred at random points, calculating the standard deviation for 
each density bin. The blue outline is the averaged distribution 
for the 21 fields centred at density peaks higher than the density 
measured at the radio galaxy position, and with $G$'s higher than 
that of the whole simulated map. 
The orange dot-dashed and the cyan dashed histograms are for the fields 
with the largest $G$ and the highest void fraction among the 21 fields, 
respectively (G435 and G400 fields). 
The brown dotted histogram is for the field around the highest-density 
peak over the whole map. 
The cumulative distributions are shown in the top panel, 
together with the 10- and 90-percentiles. 
The histograms are slightly shifted along the $x$ axis to improve 
the visibility. 
The large fractions of pixels in the highest- and the lowest-density bins 
cannot be simultaneously reproduced with the simulated density map 
within the errors. Although the high-density regions can be reproduced 
at the peak positions, the high volume fraction of the void regions 
is not seen around typical peaks. 
The G400 simulated field best reproduces both the large void fraction and 
the high density tail seen in the \tnj1338\ field. 
However the peak of the histogram around the average density 
($1+\delta_{\rm LAE}=1$) is not so high, and the high density tail is 
not so long, compared with those seen in the \tnj1338\ field. 
}
\label{fig:dhist_sim}
\end{figure}

It is also important to check the size of field-to-field variations
within the simulation volume. To do this, we first obtained  
density maps centred at random positions. 
We selected 100 positions in each of the four slices within the 
central $440\times 440\, h^{-2}\rm Mpc^{2}$, and made density 
maps within a FoV of Suprime-Cam centred at these 400 points. 
Then we calculated the standard deviation for each bin of the 
density distribution using the 400 maps. 
Fig.~\ref{fig:dhist_sim} shows that the observed density 
distribution, especially the density contrast within a single field, 
is very different from that for the simulated map as a whole. 
The volume fraction within the medium- to high-density 
($1+\delta_{\rm LAE}\sim 1-3$) bins in the \tnj1338\ field 
agrees with that obtained from the whole simulation volume, 
at $1\sigma$ level. On the other hand, the relative frequencies 
for the lowest- and the highest-density bins do not agree with 
those for the simulated map. This also suggests that such extreme 
overdensities / underdensities as we found observationally are quite 
rare in the simulation: the random sampling in the simulated map is 
unlikely to reproduce the unusual density distribution observed in our field. 
In fact, we found only 94 density peaks higher than that measured 
at the radio galaxy position, over the whole simulated map. 
Among the fields centred at the 94 peaks, only 21 has $G$'s higher than 
that of the whole simulated map (see \S\ref{sect:comp_contrast}). 
Since the \tnj1338\ field has higher $G$ 
than the whole simulation (and the SXDS field), the number density of 
\tnj1338-like field is thought to be close to that of such high-$G$ fields. 
The number density is thus comparable to that of Coma-type protoclusters 
identified in the same simulation \citep[58 protoclusters:][]{chiang2013}. 
The number (21 peaks) corresponds to number density of 
$6.4\times 10^{-8}\rm Mpc^{-3}$ in comoving units. 
This number density is also comparable to that of known bright 
radio sources at $z>2$ (see the section below). 
However, most of these peaks do not seem to reproduce the high density 
contrast seen in the \tnj1338\ field. This suggests that high-density 
regions that host powerful radio galaxies associated with a giant \lya\ 
nebulae, like our observed field, should be much rarer than this.

\subsubsection{Density fields around the high-density peaks}
\label{sect:comp_contrast}
Next we checked the density distributions around the 21 density peaks 
with large $G$'s found in the simulated map. In order to obtain the density 
distribution in the same manner as the observed data, we extracted the 
LAEs within a single Suprime-Cam FoV centred at the peak positions. 
Then we made the density maps in these fields, and calculated the 
density distributions using the grid points within the central 
$40\times 27.5\, h^{-2}\rm Mpc^{2}$, similarly to the analyses for the 
observed LAE sample. Fig.~\ref{fig:dhist_sim} also shows the average 
density distribution of the fields around the 21 density peaks, 
together with the errors. This average distribution is shifted 
toward higher density compared with the whole simulated map, 
smoothed along the density axis. However, the shape of the distribution 
does not significantly change: it agrees very closely with that of 
the whole map if we normalise the density with the average within 
these 21 fields, instead of the average of the whole simulated map. 
The distribution again seems to be different from the distribution in 
the \tnj1338\ field, although the high-density tail is reproduced well. 
This implies that high-density regions just adjacent to the void regions, 
like we observed, are still much rarer than simple high-density peaks. 
For example, we can see that the field around the density maxima 
(normalised with the average within the same field) shows much narrower 
distribution than the observed one in the \tnj1338\ field. 

To further analyse the density distribution, 
we calculated the Gini coefficient for each map centred at 
the 21 high-density peaks. Fig.~\ref{fig:hist_comp} shows the comparison 
of the Lorentz curves for the observed- and simulated density maps. 
The ordinate corresponds to the cumulative sum of the abscissa of the 
left panel, normalised with the total sum of the LAE surface density 
measured at each grid point. 
This shows that, while the density distribution of the SXDS field 
is well within the scatter of the simulated map, the distribution 
of the \tnj1338\ field is far from the simulated ones, even from those 
around most of the density peaks. 
The possible exceptions are three peaks reproducing the $G\ge 0.4$. 
These three highest-$G$ peak has $G=0.435$, 0.403, and 0.400 
(hereafter G435, G403, and G400 fields, respectively), which 
are fairly close to the observed $G$ in the \tnj1338\ field. 
Of these three, two highest-$G$ fields (G435 and G403) does not 
have sufficiently high volume fraction of the void regions, and 
have only one peak between the average- and zero density, 
as seen in Fig.~\ref{fig:dhist_sim}. The G400 field, in contrast, 
has exceptionally high void fraction ($\sim 15$ percent of the density 
cells are classified into the lowest density bin). This is the largest 
void fraction seen in the 94 peaks, which is still $\sim 1.5$ times 
higher than that in the field with the second largest void fraction 
(the second largest fraction was found in the G435 field). 
The G400 field has thus the density distribution closest to that 
of the \tnj1338\ field among the whole simulated map. 
The peak density of G400 field is $1+\delta_{\rm LAE}=3.26$ when 
normalised with the average over the whole map, and 2.86 when 
normalised within the same FoV (see Table~\ref{tab:stat_sim}). 
The other two high-$G$ fields seems to be reproducing better 
than the G400 field in terms of the peak density, but the density 
distribution (e.g., the dynamic range and the void fraction) of the 
G400 field better reproduces the properties of \tnj1338\ field. 

\begin{table*}
\centering
\begin{minipage}{160mm}
 \caption{Statistics of the simulated density field}
 \label{tab:stat_sim}
 \begin{tabular}{@{}ccccccc@{}}
 \hline
 Field & Mean $^{a}$& Peak $^{b}$ & \multicolumn{3}{c}{Percentiles $^{b}$} &
 $G$ $^c$\\
 & $[h^{2}\rm Mpc^{-2}]$ & & 10 & 50 & 90 & \\
 \hline
 G435 & 0.0409 & 3.87 (3.17) & 0.24 (0.19) & 0.71 (0.59) & 2.35 (1.92) & 0.435\\
 G403 & 0.0613 & 2.77 (3.40) & 0.24 (0.30) & 0.72 (0.89) & 2.21 (2.72) & 0.403\\
 G400 & 0.0570 & 2.86 (3.26) & 0.11 (0.13) & 0.97 (1.11) & 1.94 (2.21) & 0.400\\
 Density maxima & 0.1104 & 2.37 (5.24) & 0.48 (1.07) & 0.97 (2.15) & 1.57 (3.46) & 0.246\\
\\
 Whole map & 0.0499 & 5.19 & 0.365 & 0.92 & 1.73 & 0.301 \\
 \hline
\end{tabular}

$^a$Mean surface density over the field in comoving scale.\\
$^b$In units of the relative density $(1+\delta_{\rm LAE})$ normalised with 
the average of the FoV of Suprime-Cam in the same field. 
Parenthesises show the densities normalised with the average of the whole 
simulated map.\\
$^c$Gini coefficient of the density distribution.\\
\end{minipage}
\end{table*}

These results together suggest that the density distribution of the 
G400 is qualitatively consistent with that of the \tnj1338\ field, 
in terms of the shape of the histogram. The remaining differences 
might be explained in terms of changes in the galaxy-formation 
processes in such unusually high-density regions harbouring powerful 
radio galaxies. The simulation does not include \lya\ radiation from 
AGN activity, which can introduce significant differences in the LF, 
even if the underlying dark matter structure is similar. 
We then roughly evaluated how well the G400 field reproduces the 
\tnj1338\ field, in terms of the number density: 
we found no other G400-like fields within the whole simulation volume of 
$(500\,h^{-1}\rm Mpc)^3$ (comoving). Even the other two high-$G$ fields 
listed in Table~\ref{tab:stat_sim} do not reproduce the observed density 
distribution of the \tnj1338\ as good as the G400 field. 
Since we made four slices with thickness of $112.4\, h^{-1}\rm Mpc$, 
this corresponds to a comoving number density of 
$3.1\times 10^{-9}\rm Mpc^{-3}$, although the uncertainty 
is very large because of the small-number statistics. 
This is two orders of magnitude lower than the number density of 
Virgo-type protoclusters identified in the same Millennium Simulation, 
and still an order of magnitude lower than Coma-type protoclusters 
\citep{chiang2013}. 
The number density of radio galaxies in the universe at $2<z<5$ is 
observed to be a few $10^{-8}\rm Mpc^{-3}$ 
\citep[e.g.][]{venemans2007,miley2008}, comparable to that of the
Coma-type protoclusters. This means that the environment of the radio 
galaxy \tnj1338\ is even rarer than the known radio galaxies at $z>2$. 
In fact, not all the high-$z$ radio sources have \lya\ nebulae 
extended to $\ga 100\rm\, kpc$ \citep{vanojik1997}. Such extended 
\lya\ nebulae should reflect exceptionally active interaction between 
the radio galaxy and its surroundings. 

Note again that we found only one field like G400 (or three fields 
with $G\ge 0.4$) in the whole simulation volume, 
so the theoretically predicted number density of such 
high-density region remains significantly uncertain. 
The observational constraints are also very weak, since there are very 
few high-$z$ powerful radio galaxies known to date, and the technique 
finding high-$z$ radio sources, i.e., selecting based on ultra-steep 
radio spectra, will miss a significant fraction of high-$z$ galaxies 
\citep[e.g.][]{ker2012}. Even larger simulations together with more 
realistic models for LAEs including AGN are needed to calculate 
accurate statistics for rare density peaks like that in the \tnj1338\ 
field, and to simulate the density field and the LF found around them. 
In addition, deep and large surveys for high-$z$ radio sources, 
as well as the follow-up spectroscopy,  
are needed to more accurately estimate the observed number density 
of \tnj1338-like regions. 

\subsubsection{Density peak profiles}
\label{sect:dprofile}
\begin{figure*}
\centering
\includegraphics[width=160mm]{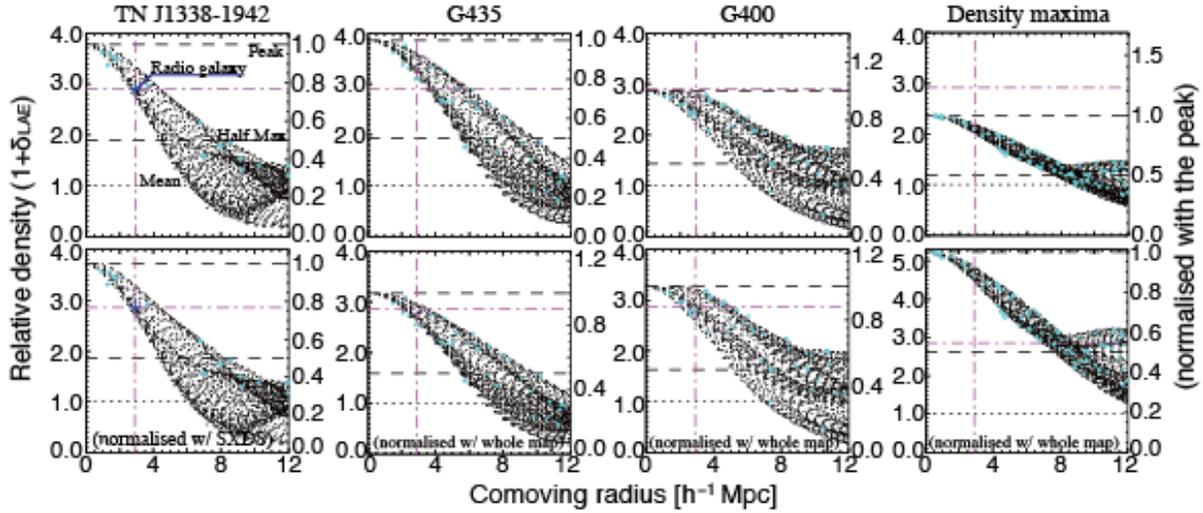}
\caption{Overdensity profiles around density peaks. 
Relative overdensity is plotted as a function of the comoving 
distance from the peak. 
The leftmost column shows the observed density profile centered 
at the density peak observed in the \tnj1338\ field. 
The profile in the top panel is based on the density normalised 
with the average over the FoV of Suprime-Cam in the same field. 
The density measured at the radio-galaxy position is shown with the 
blue cross. 
The bottom panel is based on that normalised with the average of 
the SXDS field. The remaining columns show the profiles of the 
density peaks found in the simulated map. 
For each panel, each black point corresponds to the measurement at a single 
grid point. The LAE positions are marked with cyan diamonds, with the 
symbol sizes representing the \lya\ line luminosities. 
The mean density is shown by the dotted line, and the peak- and 
the half-maximum densities are shown by the black dashed lines. 
The density and the distance measured at the position of the radio galaxy 
are marked with the magenta dot-dashed lines. 
The right axis shows the density normalised with the peak value. 
}
\label{fig:profiles}
\end{figure*}

In order to test how well the simulated map reproduces the observed 
spatial distribution of the LAE density in the \tnj1338\ field, 
we computed the radial overdensity profile of this peak, and compared 
with the simulated density map. Fig.~\ref{fig:profiles} shows the 
comparison of the density profiles between the observed peak in the 
\tnj1338\ field and high-density peaks in the simulated map. 
The leftmost panels show the observed density profile. 
As shown in \ref{sect:hdr}, the peak density in the \tnj1338\ field 
is $1+\delta_{\rm LAE}=3.80$, and the radio galaxy position is slightly 
offset ($\sim 2.8\, h^{-1}$ comoving Mpc) from the peak, at which the 
density is $\approx 77$ percent of the peak. 
Although the redshift uncertainty is fairly large, we can roughly 
estimate the spatial scale wherein the high-density peak is 
accumulating the material from the surroundings, based on this profile. 
The radius of half-maximum is $\sim 6\, h^{-1}$ projected comoving Mpc, 
and the average density within this radius is $1+\delta_{\rm LAE}\sim 3$. 
This could be produced by spherical collapse from a region with 
radius of $\sim 9\, h^{-1}\rm Mpc$ (comoving). Although this is a 
very rough estimate, it is clear that the accumulation of material 
is occurring well within our survey volume. The profile does not 
significantly change even when normalised with the average density 
over the SXDS field.

For comparison with this observed density peak profile, we analysed 
the 94 high-density fields described above (including the three 
highest-$G$ fields). 
We found that only one of these fields (G435) has peak density as high as 
the observed one in \tnj1338\ field ($1+\delta_{\rm LAE}=3.80$), and 
11 fields have the peak density higher than the observed density at 
the radio galaxy position ($1+\delta_{\rm LAE}=2.91$), when we normalise 
the density with the average {\em within} the same Suprime-Cam-sized FoV. 
When we look into the peak densities normalised with the average 
over the whole simulated map, 11 have their peak values higher than 
the observed one in the \tnj1338\ field. All except one with $G=0.356$ 
have $G$'s smaller than 0.3. These together suggest that not all the 
peaks as high as the observed one show high dynamic range within 
$\sim 50\, h^{-1}$ comoving Mpc scale. To test how the simulated maps 
are reproducing such dynamic range, we analysed the radial density 
profiles of the simulated density peaks. 
Fig.~\ref{fig:profiles} compares the profiles of three 
simulated peaks with the observed profile. Table~\ref{tab:stat_sim} 
summarises the statistics of the simulated peaks shown in 
Fig.~\ref{fig:profiles}. 
The three simulated fields shown here are those with the highest $G$ 
(G435), the largest void fraction (G400: this field also has the third 
highest $G$), and the highest peak density (density maxima when normalised 
with the whole map) among all the 94 peaks identified above. 

The high peak density as we found in the \tnj1338\ field is rarely 
seen in the simulated peaks when normalised within the same FoV. 
The exceptional cases are only three out of the 94, including the 
G435 field. The average density of the G435 field is $\sim 0.8$ 
times the whole map, which is the lowest average density of 
all the fields centred at the 94 peaks. Although the volume fraction 
of the void regions is not so large in this field as in the \tnj1338\ 
field, the radial density profile of this peak seems to be reproducing 
the observed one fairly well. 
Most of the fields around the 94 peaks, including the G400 field, have 
their average densities higher than the whole map, in contrast. 
Even for the G400 field, the average density is higher than the whole 
simulated map by a factor of $\sim 1.1$. We can see that the peak 
density is significantly lowered when normalised within the same FoV, 
although the peak width is comparable to that of the observed peak 
in the \tnj1338\ field. 
This difference is still much larger in the 
field around the density maxima. The maximum density exceeds 
$1+\delta_{\rm LAE}=5$ when normalised with the average of the whole map, 
but goes down to $\sim 2.4$ when normalised within the same FoV. 
The density contrast in this field is thus fairly small compared 
with the \tnj1338\ field. 

Except for the field around the density maxima, wherein the central 
peak has fairly extended outskirt, the shape of the observed density 
profile seems to be roughly reproduced by the remaining two simulated 
peaks selected here. 
The half-maximum radii are $\sim 6\, h^{-1}\rm Mpc$, and the peak 
densities agree with that we observed in the \tnj1338\ field 
within the error (within $\pm 0.5$, although it depends on the 
normalisation). The Gini coefficients are also large, $G\ge 0.4$. 
Comparing the two cases, G435 and G400, 
the G435 shows better match with the observed profile of the \tnj1338\ 
field in terms of the amplitude and the width of the peak. 
G400 field also seems to be reproducing the observed profile, 
but the peak density is only $\sim 3$ when normalised within the 
same FoV. The shape of the peak seems slightly elongated to form 
small plateau, leading to the extended component of the peak profile. 
The peak in the \tnj1338\ field seems to be more isolated, 
while in the G400 field the main peak is facing next to the minor 
peak with $1+\delta_{\rm LAE}\sim 2$ leading to the relatively broad 
outskirt. 
These results together shows that none of the simulated fields 
simultaneously reproduce the sharp peak profile and the broad density 
distribution around the peak as we found in the \tnj1338\ field. 

There are at least three factors which might explain the difficulties 
in reproducing the density distribution observed in the \tnj1338\ field. 
First, the selection of LAEs in the simulation does not exactly
match the observational one. Especially for the broadband colours, 
we did not put any constraints to select the LAEs from the mock 
catalogue, possibly causing the difference in the selection function. 
However, this effect is unlikely to significantly change the resulting 
density distribution. Our mock LAE sample reproduces both the LF and 
the density distribution of the SXDS control sample quite well. 
Our selection of LAEs is thus thought to be working well at least 
for the faint end of the LF, where the \tnj1338\ and the mock 
samples agree with each other. 
The second possibility is that the LAE model itself is not 
sufficiently realistic. For example, the escape fraction of \lya\ 
photons is assumed to be constant for all the LAEs. If the escape 
fraction is changed in highly overdense regions by e.g. galaxy-galaxy 
interactions or AGN feedback, the spatial distribution of LAEs may 
significantly change in the field. 
This is also related to the third factor, i.e., the observations might 
include a significant contribution of \lya\ coming from AGN, which is 
not included in the model. AGN activity may affect the galaxy-formation 
processes of the surroundings through, e.g., induced star formation 
\citep{zirm2005}. 
Such AGN activity may be enhanced in overdense environments 
\citep[e.g.][]{pentericci2002,croft2005,lehmer2009,digby-north2010}. 
If the bright LAEs in our \tnj1338\ sample are predominantly AGNs, 
this could lead to significant enhancement of the LAE number density, 
emphasizing the density contrast. 

Another, much more naive interpretation is that the 
Millennium Simulation is not large enough to reproduce this kind 
of overdensity. This is in case if the halo mass of the overdensity 
is significantly larger than the maximum halo mass of the Millennium 
Simulation. Even if the SXDS field is suffering from overdensities 
and our density normalisation is not correct, such a steep and high 
density peak with large density contrast, as well as the shape of the 
density histogram, is not reproduced in the current simulation. 
The disagreement may be overcome by carrying out even larger simulations. 
The density profiles around massive haloes in such a large simulation may 
be different from those around less massive haloes: extremely 
massive haloes possibly form very high density peaks without 
significant growth in size of the overdense-regions. 

The observational uncertainties may also affect the density distribution. 
Since we do not have accurate redshifts for most of our LAEs, 
our estimate of the \lya\ luminosity have large uncertainty. 
This may especially affect the completeness for the faint sources, 
and our detection completeness at faint end may be lower than expected. 
The density distribution of the mock LAEs is dominated by relatively 
faint LAEs, so that some fraction of faint LAEs might be missed in 
our observed data, resulting in apparently high peak density and 
density contrast. Indeed, when we use 30 brightest LAEs within each 
field of the simulation to compute the density map, instead of 
normalising the density, the density distribution for some fields 
in the simulation becomes closer to the observed one in the 
\tnj1338\ field. However, even for the field giving the best match 
with the observed density distribution in this case, such a 
steep, isolated density peak cannot be reproduced. We still need 
fine tuning to reproduce the simulated density field with observed one. 

In any case, more precise measurement of the density field and 
constraints on the AGN activity in our LAEs are needed to 
test the possibilities described here.

\subsubsection{Luminosity function}
\label{sect:comp_lf}
Fig.~\ref{fig:lf_obs} also shows the LF of the mock LAE sample. 
Again this sample is selected with the same colour constraints 
as the \tnj1338\ sample. The mock cagalogue has been already shown 
to have a \lya\ LF reproducing well that of O08's sample 
\citep{orsi2008}. This means that the mock sample is expected to 
reproduce the LF of our control sample. 
The LF of the mock sample agrees quite well with the control 
sample, although except for the bright end ($\log L_{\rm Ly\alpha}\sim 43.6$). 
This difference is thought to come from the cosmic variance. 
The bright end of the LF of our control sample is determined with 
a single source, and the probability of detecting such bright sources 
within a single FoV is relatively low. Even for the fields around 
the high-density peaks with large $G$'s selected above, 
10 out of the 21 fields do not contain any sources with 
$\log L_{\rm Ly\alpha}\sim 43.6$ 
(the remaining 11 fields conatin only one such bright source each). 

Similarly to the comparison with the blank fields, 
the LF in the whole \tnj1338\ field is higher than the 
mock LAE LF at the bright end, but the difference 
is much larger than the case of the control sample. 
The LF of the whole \tnj1338\ field is factor of $\sim 10$ 
higher than the mock LF at the bright end, $\log L_{\rm Ly\alpha}\sim 43.6$. 
The difference becomes a factor of $\sim 20$ when we look 
into the LF of the overdense regions of \tnj1338\ field. 
For the fainter range, the \tnj1338\ LF comes below that of 
mock LAEs at $\log L_{\rm Ly\alpha}\sim 43.0$. 
The crossing point is in between these two, i.e., 
the number of LAEs fainter than $\log L_{\rm Ly\alpha}\sim 43.0-43.3$ 
seems to be relatively suppressed in this field. 
The faint end of the LF agrees with the mock, when we look at 
the LF of the overdense regions in the \tnj1338\ field. 
While the faint-end LF of this field does not show any enhancement 
compared to the mock LF, the bright end shows a large enhancement 
by one order of magnitude. 
This again suggests that the \tnj1338\ field is a peculiar 
field dominated by bright LAEs, or AGN hosts including the host of 
the radio galaxy. 
There are four bright LAEs contained in the two highest-luminosity 
bins ($\log L_{\rm Ly\alpha}\sim 43.6$ and 43.9) in this field, 
and such bright sources are likely to be hosting AGN, as suggested by O08. 
Such a population of bright LAEs, as well as the unusually high 
galaxy overdensity, should largely affect the galaxy-formation 
within this field. 

The main differences between the \tnj1338\ and the mock LAE samples 
in terms of LFs are then thought to be 
(1) the presence of the AGN, 
(2) the high-density peak as high as $1+\delta_{\rm LAE}\sim 4.1$, and 
(3) the large density contrast within a single FoV of the Suprime-Cam. 
Of these three, (2) can be partly reproduced with the mock sample as 
described in \S\ref{sect:comp_whole} and \S\ref{sect:dprofile}. 
We then tested how the LF differs from the observed one when one 
of these three main discrepancies are partly solved, 
by computing the LF within the peak regions seen in the simulated map. 
The resulting LFs in overdense regions show almost the same 
shape as the LF from the whole mock sample, and the 
amplitude of the LF becomes higher by up to 0.4 dex. 
We can see in  Fig.~\ref{fig:lf_obs} that only $\sim 50\%$ (11 fields) 
of the 21 high-density fields contain LAEs with 
$\log L_{\rm Ly\alpha}\sim 43.6$. 
Although most of these 11 fields contain only one source each, 
we can estimate the bright end of the LF by averaging the 
21 fields, $\log n_{\rm LAE} = -5.56$ at the bright end. 
This value is only 0.08 dex higher value than the total mock sample. 
This means that there are no significant enhancements of the number 
(fraction) of bright LAEs such as we found in the \tnj1338\ field. 
While the enhancements of the total number density of LAEs are 
relatively well reproduced in these high-density fields, the LF shape 
remains almost unchanged from that of the whole mock sample, 
i.e., not biased to bright sources. 

Note that not all of the high-density peaks described in 
\S\ref{sect:dprofile} are included in the 11 fields 
containing such brightest LAEs: the field around density maxima contains 
only one, and G435, G403 and G400 fields contain no such LAEs as 
bright as $\log(L_{\rm Ly\alpha})\sim 43.6$. 
This simply means that the brightest LAEs are 
likely to be found in fields with high average density, 
but not necessarily with high density contrast. 
This is also expected from the properties of the mock LAEs: 
the galaxy bias positively correlates with the \lya\ luminosity 
$L_{\rm Ly\alpha}$, but changes by only a factor of $\sim 2-3$ 
\citep{ledelliou2006,orsi2008}. On the other hand, the LFs of 
our {\em observed} samples suggest that the bias depends more 
strongly on  $L_{\rm Ly\alpha}$. The mock LAE sample does not contain 
AGN at all, so that it can reproduce only the properties of 
``normal'' LAEs, not LAEs hosting AGNs, nor AGN-induced star formation. 
Activity related to such AGN, e.g., AGN-induced star formation 
in the vicinity of the radio galaxy \citep{zirm2005}, 
may lead to enhancements of star-formation in this high-density region. 
Such interactions between the AGN and the surrounding 
environment may also occur further out from the central radio 
galaxy, as the extended \lya\ nebula is observed up to 
$\sim 100\, \rm kpc$ \citep{venemans2002}. 
For some \lya\ nebulae, interaction is (although indirectly) suggested 
even further out from their alignment with the surrounding large-scale 
structure up to $\sim 10\, h^{-1}$ comoving Mpc \citep{erb2011}. 
Furthermore, three out of the four LAEs mentioned above in 
the highest luminosity bins in the \tnj1338\ sample, as well as the 
radio galaxy, lie within the overdense ($1+\delta_{\rm LAE}> 1$) regions. 
These may together suggest that star-formation / AGN activity 
is enhanced by the presence of AGN in the overdense regions, 
while there are no such sources of enhancement in the underdense regions. 

Such an enhancement is expected from the high galaxy density, 
leading to frequent galaxy mergers. The high density may also 
lead to high gas accretion rates, which enhances the star-formation / AGN 
activity resulting in bright LAEs. These mechanisms will result in 
a large bias of the bright LAEs, dominating the bright end of the LF. 
Such a large bias for bright LAEs results in the large density contrast 
of bright LAEs described in \S\ref{sect:d_contrast}. 
In order to see how such an enhancement affects the LF of {\em normal} 
star-forming galaxies in the overdense environment, we need to take 
deeper images to construct a sample of much fainter LAEs. 
We still have difficulties to make a meaningful comparison with 
the simulation, due to the shallowness of the data: the luminosity 
range of our observed samples corresponds to the very bright end 
of the mock LAEs. Constructing a sample of fainter LAEs 
(down to $\log(L_{\rm Ly\alpha})\sim 42.7$ or fainter), 
which is less likely to be contaminated by AGN, is essential to probe 
the LF and its environmental dependence. The uncertainty in the 
redshifts is also a problem in computing the galaxy density and the 
luminosities, as well as evaluating the contamination of AGN and 
foreground galaxies. Further deep imaging to construct a deeper sample, 
and subsequent follow-up spectroscopy, are needed to draw 
further information from this comparison.

\section{Conclusions}
We have carried out intermediate-band and broadband imaging 
observations with Suprime-Cam on the Subaru Telescope of a $z=4.1$ 
radio galaxy associated with a giant \lya\ nebula to probe its 
environment on a $\sim 50\, h^{-1}\rm Mpc$ scale. In order to quantify 
the environment with a control sample, we utilized the existing data 
from a blank field, SXDS, taken with the same instrument and 
similar filters. We also used a mock LAE catalogue generated with a 
semi-analytical galaxy formation model based on the Millennium Simulation. 

We confirmed that the radio galaxy lies in a region with peak 
LAE number density of $1+\delta_{\rm LAE}= 3.8\pm 0.5$ at the 
radio galaxy position, even after taking account of the relatively 
large redshift range covered by the {\it IA624} filter. 
The average density over the Suprime-Cam FoV is almost the same 
as that of the blank field within 2 percent, showing that the 
density contrast in this field is very high. These results suggest 
that radio galaxies associated with \lya\ nebulae emerge in extreme 
overdensity environments. Comparing these with the mock LAEs, 
we found that such high-density regions are quite rare, 27 peaks 
in the whole simulation volume of $(500\, h^{-1}\rm Mpc)^3$, 
and correspond to the densest $< 0.4$ percentile. 
The corresponding number density is comparable to that of Coma-type 
protoclusters found in the same simulation, and known radio 
galaxies at $z>2$. 
The overdensity associated with the radio galaxy can be traced 
up to $\sim 3-6$ comoving Mpc, facing the large void 
region just adjacent to it. Only one of the 35 density peaks 
have such a large density contrast on this scale, and even this 
peak cannot fully reproduce the density peak profile. 

We found the density dependence of the luminosity functions (LFs). 
The \lya\ LF of the \tnj1338\ field shows enhancement at the bright 
end ($\log(L_{\rm Ly\alpha}[\ergs])\ga 43.3$) by an order of magnitude 
or more, while the faint end almost agrees with that in the blank field. 
We compared the LFs with the mock LAEs to conclude that 
star-formation and/or AGN activities affecting the bright end 
of the LF are highly enhanced within overdense regions. 
We pointed out the possibilities that frequent galaxy mergers or 
high gas-accretion rate enhance the star-formation / AGN activity, 
or AGN-induced star formation in this field. 
The presence of a powerful radio galaxy showing a giant \lya\ 
nebula is thus a possible signpost of changes in galaxy formation 
activity on scales of $\sim 3-6$ comoving Mpc.

\section*{Acknowledgements}
We achnowledge the anonymous referee, who gave us important and 
practically useful comments to improve the draft. 
We thank Masami Ouchi for providing the data of luminosity functions, 
and giving us useful comments. 
We thank Yi-Kuan Chiang, Roderik Overzier, Surhud More, John Silverman, 
Masao Mori, and Takatoshi Shibuya for having discussion over this work, 
and giving us valuable comments. 
This work was supported by the FIRST programme ``Subaru Measurements 
of Images and Redshifts (SuMIRe)'', World Premier International 
Research Center Initiative (WPI Initiative), MEXT, Japan. 
This work was supported in part by the Science and Technology 
Facilities Council rolling grant ST/I001166/1 to the ICC.
Calculations were partly performed on the ICC Cosmology Machine,
which is part of the DiRAC Facility jointly funded by STFC and 
Durham University. 
YM acknowledges support from JSPS KAKENHI Grant Number 20647268.
IRS acknowledges support from STFC (ST/I001573/1), 
the ERC Advanced Investigator programme DUSTYGAL 321334 and a 
Royal Society/Wolfson Merit Award.
The data were in part obtained from SMOKA, which is operated 
by the Astronomy Data Center, National Astronomical Observatory of Japan.

\appendix
\section{Three-dimensional density profiles}
As discussed in the \S\ref{sect:comp_contrast}, 
such a high density peak and a large void region within a single 
FoV as we found in the \tnj1338\ field is likely to be reproduced 
by fields with large $G$'s, like the G435 or G400 field. 
However, although the G435 field has very large $G$ 
the volume fraction of the void regions is not so high as that 
in the \tnj1338\ field, the shape of the peak profile is relatively 
well reproduced. The G400 field reproduces the observed large void 
fraction, the shape of the density histogram, and the small difference 
between the two cases of normalisation, but does not reproduce the 
peak profile well compared with the G435 field. 
Unlike these high-$G$ fields, the field around the density maxima 
have rather different peak profile and density distribution, although 
the peak density is the highest when normalised with the whole map. 

In addition to such statistics of the two-dimensional density 
distribution, the three-dimensional (3-D) density gives useful 
information to infer how well the simulation reproduces the 
observed density field, as well as to infer the underlying 
3-D structure of the \tnj1338\ field. 
We calculated the local- and global densities using the 3-D 
spatial distribution of the mock LAEs within these fields. 
The density at the position of each source 
was calculated based on the distance to the 5th-, 10th- and 20th-nearest 
neighbour. We normalised the density with the average number density over 
the same field, and calculated the overdensity $\delta_{\rm 5th}$, 
$\delta_{\rm 10th}$, and $\delta_{\rm 20th}$. 
Then we counted the LAEs residing in the density greater than 
the average. Table~\ref{tab:density} summarises the results of 
the density measurements of these three high-density fields, 
comparing with the observed data. 

\begin{table*}
\centering
\begin{minipage}{160mm}
  \caption{Density measurements around the high-density peaks}
  \label{tab:density}
  \begin{tabular}{@{}lcccccc}
  \hline
 Field & $N_{\rm tot}$$^a$ & ${\rm max}(1+\delta_{\rm LAE})$$^b$ 
 & $N(\delta_{\rm LAE}>0)$$^c$ 
 & $N(\delta_{\rm 5th}>0)$$^d$ & $N(\delta_{\rm 10th}>0)$$^d$ 
 & $N(\delta_{\rm 20th}>0)$$^d$\\
\hline
 G435 & 66 & 3.9 & 39 (34)& 16 & 10 & 1\\
 G403 & 112 & 2.8 & 74 (84)& 38 & 27 & 2\\
 G400 & 92 & 2.9 & 69 (73)& 27 & 6 & 0\\
 Density maxima & 192 & 2.4 & 121 (174)& 74 & 53 & 43\\
\hline
 TN J1338-1942 & 30$^e$ & 3.8 & 26$^e$ & - & - & -\\
 (SXDS)$^f$ & 34 & 2.4 & 29 & - & - & -\\
\hline
\end{tabular}

$^a$Total number of LAEs within the field.\\
$^b$The peak (maximum) surface density within the field.\\
$^c$The number of LAEs with the surface density greater than the average. 
The numbers in parenthesises are based on the average over the whole map.\\
$^d$The number of LAEs with the 3-D densities greater than the average over 
the whole simulation volume. The densities were calculated from the 5th-, 
10th- and 20th-nearest neighbours for $\delta_{\rm 5th}$, 
$\delta_{\rm 10th}$ and $\delta_{\rm 20th}$, respectively.\\
$^e$The radio galaxy is not counted.\\
$^f$Control field (not a high-density peak).\\
\end{minipage}
\end{table*}

\begin{figure*}
\includegraphics[width=160mm]{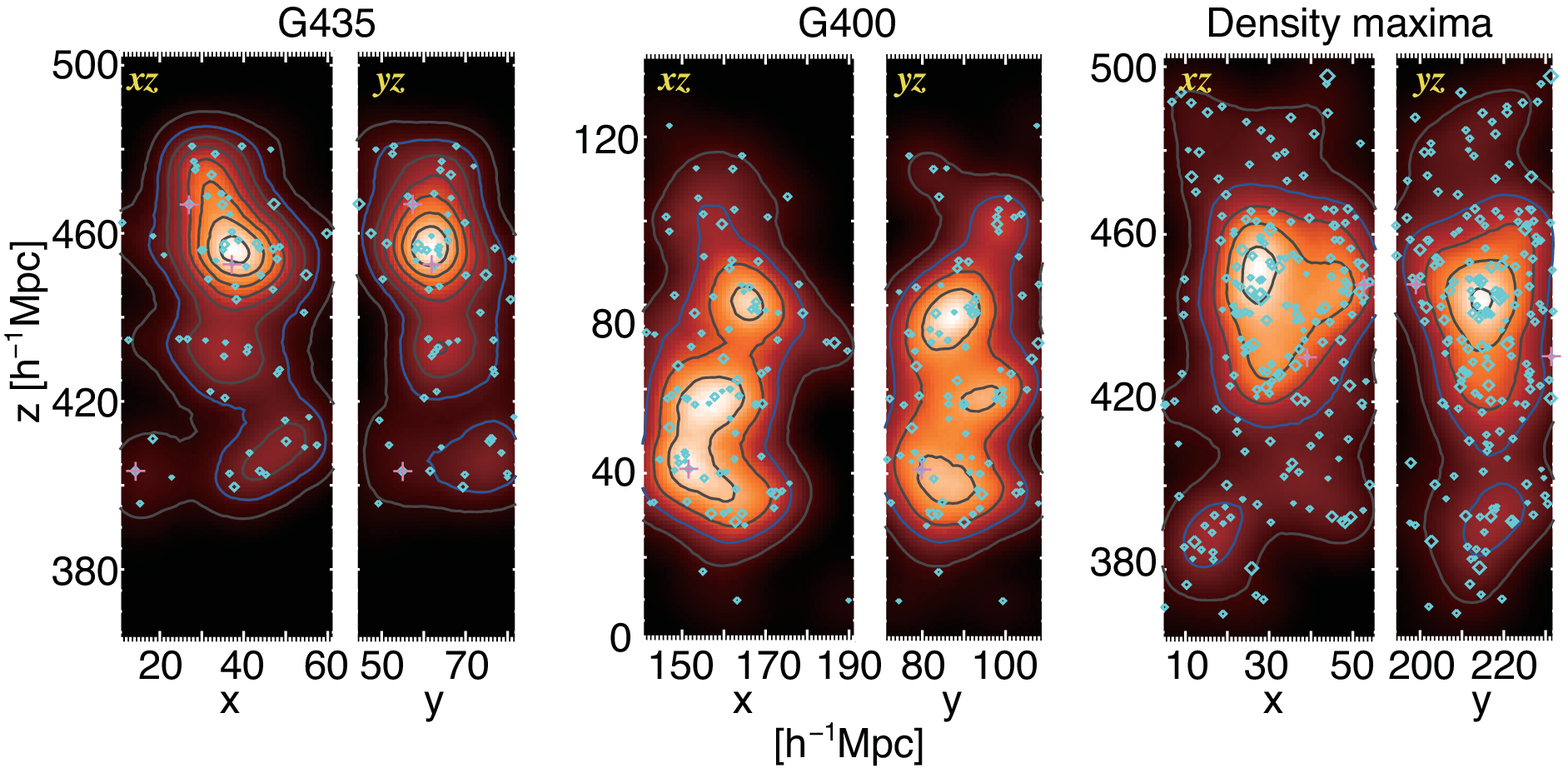}
\caption{LAE density maps of the three fields with high-density peaks, 
generated with the mock LAEs projected onto the $xz$- and $yz$ planes, 
instead of the $xy$ plane shown in Fig.~\ref{fig:dmap_comp}. 
The leftmost two panels are the maps for the G435 field, projected on the 
$xz$- (left) and the $yz$ plane (right). The two panels on the centre 
are for the G400 field, and the rightmost two panels are for the 
field centred at the density maxima. 
All the coordinates shown here are in the comoving scales. 
The maps are drawn with the same manner as in Fig.~\ref{fig:dmap_comp}, 
except for the use of smoothing radius of $8\, h^{-1}\rm Mpc$. 
The density shown here is normalised with the average over the 
same field. LAEs associated with the largest halo mass over each 
field are marked with the magenta crosses. 
For all the three peaks shown here, there seems to be elongated filamentary 
structure nearly along the $z$ axis. This is quite common in the 94 
high-density peaks we found in the simulated density map. 
The overdense region in the G435 field emerged on the $xy$-plane density 
seems to be dominated by a single hidh-density peak at $z\sim 460$. 
This peak is relatively compact compared with that seen in the density 
maxima around $z\sim 440$. For the G400 field, in contrast, there are 
at least three compact high-density clumps nearly along the $z$ axis. 
Such alignment of subclimps can be seen in most of the 21 high-$G$ 
peaks we found in the simulated map. 
}
\label{fig:dmap_xz+yz}
\end{figure*}

Comparing the three high-$G$ fields and the field around the density 
maxima, we cannot see significant differences in the fraction of LAEs 
lying within the overdensities in terms of $\delta_{\rm 5th}$. 
On the other hand, the fraction of LAEs lying within the $\delta_{\rm 20th}>0$ 
is much higher around the density maxima than in the other three high-$G$ 
fields ($\sim 20$ percent for the maxima, and $<2$ percent for the three 
high-$G$ fields). This means that the clustering signal of LAEs is 
comparable for all the four high-density fields in $\sim 14-17\,h^{-1}$ 
Mpc scale, but much stronger around the density maxima in $\sim 22\, h^{-1}$ 
Mpc scale (both in comoving units). The overdensity (i.e., protocluster) 
associated with the density maxima is thus thought to be more evolved 
compared with other three high-$G$ fields, resulting in such a large 
size filling a large fraction of our FoV to have relatively low $G$. 
Fig.~\ref{fig:dmap_xz+yz} clearly shows such trend. 
For high-$G$ fields, size of each overdense clump is relatively small, 
and the fraction of void regions is higher than in the density maxima. 
The surface-density peak in the G435 field seems to be dominated by 
a single density peak at $z\sim 460$, but we can see at least three peaks 
along the $z$ axis in the G400 field (the G403 field similarly has three 
or more peaks nearly along the $z$ axis, although not shown in 
Fig.~\ref{fig:dmap_xz+yz}).

As seen in Fig.~\ref{fig:dmap_xz+yz}, most of the high-density peaks 
are likely to be an elongated structures or two or more small high-density 
clumps aligned with the line of sight, rather than single extremely 
overdense regions. 
If radio galaxies are formed within single extremely overdense regions, 
the high-density region seen around the density maxima is more likely 
to be reproducing the environment of the radio galaxy. 
However, as expected from the lower $G$ compared with the observed one 
in the \tnj1338\ field, the field around the density maxima does not 
have such a large fraction of void regions. The G400 field, which has 
an exceptionally large void fraction, has at least three subclumps 
along the filamentary structure. Since this is quite common feature 
in high-density fields (or in high-$G$ fields), the overdensity found 
in the \tnj1338\ field might be the result of such alignment of more 
than two subclumps. The observed overdensity around the radio galaxy 
might be dominated by a single extremely overdense clump as well, 
as the G435 field reproduces the observed density profile better than 
the G400 field (see Fig.~\ref{fig:profiles}). 

We also analised the relation between the 3-D density distribution 
and the maximum dark-halo mass hosting the LAEs in the fields around 
the 94 peaks. We found five peaks containing a dark halo with the 
maximum mass in the simulation, $9.4\times 10^{12}M_\odot$. 
These five peaks all have relatively low gini coefficients, $G\le 0.27$, 
and none of the four peaks listed in Table~\ref{tab:density} are 
included in these five. The most massive haloes found in the four 
peaks in Table~\ref{tab:density} have the mass of $2.4\times 10^{12}M_\odot$, 
which are contained in the G400 and G403 fields. 
These massive haloes are apparently associated with the highest 
density peak of each field. The maximum halo mass found in 
the remaining two fields are $8.3\times 10^{11}M_\odot$ and 
$1.7\times 10^{12}M_\odot$ for the G435 and the density maxima, 
respectively. For the G435 field one of the most massive dark halo 
can be seen very close to the highest density peak, 
while for the density maxima such massive haloes can be seen only 
at the outskirts of the peak. 
There seems to be no significant correlations between 
the maximum halo mass of the field and the density peak profiles. 
On the other hand, the peak of the density maxima does not have 
the most massive dark haloes around its centre, while the three 
high-$G$ fields all have the most massive haloes close to the centres 
of their highest peaks. This difference might be reflecting how 
the overdensities have accreted their mass until the observed epoch. 

These analyses together suggest that the simulation cannot 
fully reproduce the density distribution of the \tnj1338\ field. 
It is likely that more than two (relatively compact) high-density 
clumps are well aligned with the line of sight, at least one of which 
have the most massive dark haloes of the system near the centres. 
In any case, further detailed analyses with more realistic models 
of galaxy formation are needed to draw definite conclusions on the 
true 3-D distribution of LAEs and the mass-accretion mechanisms of 
the protocluster. 

\label{lastpage}


\begin{thebibliography}{99}
\bibitem[Atek et al.(2011)]{atek2011}
Atek H., et al. 2011, ApJ, 743, 121
\bibitem[Baba et al.(2002)]{baba2002}
Baba H., et al. 2002, ADASS XI eds. D. A. Bohlender, D. Durand, and 
T. H. Handley, ASP Conference series, 281, 298
\bibitem[Basu-Zych \& Scharf(2004)]{basu-zych2004}
Basu-Zych A., Scharf C., 2004, ApJ, 615, L88
\bibitem[Bertin \& Arnouts(1996)]{bertin1996}
Bertin E., Arnouts S., 1996, A\&AS, 117, 303
\bibitem[Blain et al. (2004)]{blain2004}
Blain A. W., Chapman S. C., Smail I., Ivison R., 2004, ApJ, 611, 725
\bibitem[Bruzual \& Charlot(2003)]{bc03}
Bruzual G., Charlot S., 2003, MNRAS, 344, 1000
\bibitem[Capak et al.(2011)]{capak2011}
Capak P. L., et al., 2011, Nature, 470, 233
\bibitem[Chambers et al.(1990)]{chambers1990}
Chambers K. C., Miley G. K., van Breugel W. J. M., 1990, ApJ, 363, 21
\bibitem[Chiang et al.(2013)]{chiang2013}
Chiang Y.-K., Overzier R., Gebhardt K., 2013, ApJ, 779, 127
\bibitem[Colbert et al.(2011)]{colbert2011}
Colbert J. W., Scarlata C., Teplitz H., Francis P., Palunas P., 
Williger G. M., Woodgate B., 2011, ApJ, 728, 59
\bibitem[Cole et al.(2000)]{cole2000}
Cole S., Lacey C. G., Baugh C. M., Frenk C. S., 2000, MNRAS, 391, 168
\bibitem[Croft et al.(2005)]{croft2005}
Croft S., Kurk J., van Breugel W., Stanford S. A., de Vries W., Pentericci L., 
Ro\" ttgering H., 2005, AJ, 130, 867
\bibitem[Dawson et al.(2004)]{dawson2004}
Dawson S. et al., 2004, ApJ, 617, 707
\bibitem[Dawson et al.(2007)]{dawson2007}
Dawson S., Rhoads J. E., Malhotra S., Stern D., Wang J.-X.,  
Dey A., Spinrad H., Jannuzi B. T., 2008, ApJ, 671, 1227
\bibitem[De Breuck et al.(1999)]{debreuck1999}
De Breuck C., van Breugel W., Minniti T., Miley G. K., 
R\" ottgering H. J. A., Stanford S. A., Carilli C., 1999, A\&A, 352, L51
\bibitem[De Breuck et al.(2001)]{debreuck2001}
De Breuck C., et al., 2001, ApJ, 121, 1241
\bibitem[De Breuck et al.(2004)]{debreuck2004}
De Breuck C. et al., 2004, A\&A, 424, 1
\bibitem[Dekel et al.(2009)]{dekel2009}
Dekel A., et al., 2009, Nature, 457, 22
\bibitem[Digby-North et al.(2010)]{digby-north2010}
Digby-North J. A., et al., 2010, MNRAS, 407, 853
\bibitem[Dijkstra et al.(2006)]{dijkstra2006}
Dijkstra M., Haiman Z., Spaans M., 2006, ApJ, 649, 14
\bibitem[Dijkstra \& Loeb(2009)]{dijkstra2009}
Dijkstra M., Loeb A., 2009, MNRAS, 408, 1109
\bibitem[Dijkstra \& Kramer(2012)]{dijkstra2012}
Dijkstra M., Kramer R., 2012, MNRAS, 424, 1672
\bibitem[Erb et al.(2011)]{erb2011}
Erb D. K., Bogosavljevi\' c M., Steidel C. C., 2011, ApJ, 740, L31
\bibitem[Fardal et al.(2001)]{fardal2001}
Fardal M. A., Katz N., Gardner J. P., Hernquist L., Weinberg D. G., 
Dav\' e R., 2001, ApJ, 562, 605
\bibitem[Faucher-Gigu\' ere et al.(2010)]{faucher-giguere2010}
Faucher-Gigu\' ere C.-A., Kere\v s D., Dijkstra M., Hernquist L., 
Zaldarriaga M., 2010, ApJ, 725, 633
\bibitem[Fukugita et al.(1995)]{fukugita1995}
Fukugita M., Shimasaku K., Ichikawa T., 1995, PASP, 107, 945
\bibitem[Furlanetto et al.(2005)]{furlanetto2005}
Furlanetto S. R., Schaye J., Springel V., Hernquist L., 2005, ApJ, 622, 7
\bibitem[Furusawa et al.(2008)]{furusawa2008}
Furusawa H. et al., 2006, ApJS, 176, 1
\bibitem[Geach et al.(2009)]{geach2009}
Geach J. E., Alexander D. M., Lehmer B. D., Smail I., Matsuda Y., 
Chapman S., Sharf C. A., Ivison R. J., Volonteri M., Yamada T., 
Blain A. W., Bower R. G., Bauer F. E., Basu-Zych A., 2009, 
ApJ, 700, 1
\bibitem[Geach et al. (2005)]{geach2005}
Geach J. E., Matsuda Y., Smail I., Chapman S. C., Yamada T., 
Ivison R. J., Hayashino T., Ohta K., Shioya Y., Taniguchi Y., 2005, 
MNRAS, 363, 1398
\bibitem[Goerdt et al.(2010)]{goerdt2010}
Goerdt T., Dekel A., Sternberg A., Ceverino D., Teyssier R., 
Primack J. R., 2010, MNRAS, 407, 613
\bibitem[Hayashino et al.(2000)]{hayashino2000}
Hayashino T., et al., 2000, Proc. SPIE, 4008, 397
\bibitem[Intema et al.(2006)]{intema2006} 
Intema H.~T., Venemans B.~P., Kurk J.~D., Ouchi M., Kodama T., 
R{\"o}ttgering H.~J.~A., Miley G.~K., Overzier R.~A., 2006, A\&A, 
456, 433 
\bibitem[Ivison et al.(2013)]{ivison2013}
Ivison R. J., et al., 2013, ApJ, 772, 137
\bibitem[Iye et al.(2004)]{iye2004}
Iye M., et al., 2004, PASJ, 56, 381
\bibitem[Kakazu et al.(2007)]{kakazu2007}
Kakazu Y., Cowie L., Hu E., 2007, ApJ, 668, 853
\bibitem[Kennicut(1983)]{kennicut1983}
Kennicut R. C., 1983, MNRAS, 301, 569
\bibitem[Ker et al.(2012)]{ker2012}
Ker L. M., Best P. N., Rigby E. E., R\" ottgering H. J. A., Gendre M. A., 
2012, MNRAS, 420, 2644
\bibitem[Kere\v s et al.(2005)]{keres2005}
Kere\v s D., Katz N., Weinberg D. H., Dav\' e, R., 2005, MNRAS, 363, 2
\bibitem[Kova\v c et al.(2007)]{kovac2007}
Kova\v c K., Somerville R. S., Rhoads J. E., Malhotra S., Wang J.-X., 
2007, ApJ, 668, 15
\bibitem[Koyama et al.(2013)]{koyama2013}
Koyama Y., et al., 2013, MNRAS, 434, 423
\bibitem[Le Delliou et al.(2005)]{ledelliou2005}
Le Delliou M., Lacey C., Baugh C. M., Guiderdoni B., Bacon R., 
Courtois H., Sousbie T., Morris S. L., 2005, MNRAS, 357, L11
\bibitem[Le Delliou et al.(2006)]{ledelliou2006}
Le Delliou M., Lacey C., Baugh C. M., Morris S. L, 2006, MNRAS, 365, L712
\bibitem[Lehmer et al.(2009)]{lehmer2009}
Lehmer B. D., et al., 2009, MNRAS, 400, 299
\bibitem[Madau(1995)]{madau1995} Madau P., 1995, ApJ, 441, 18
\bibitem[Massey et al.(1988)]{massey1988} 
Massey P., Strobel K., Barnes J.~V., Anderson E., 1988, ApJ, 328, 315 
\bibitem[Matsuda et al.(2006)]{matsuda2006}
Matsuda Y., Yamada T., Hayashino T., Yamauchi R., Nakamura Y., 
2006, ApJ, 640, L123
\bibitem[Matsuda et al.(2011)]{matsuda2011}
Matsuda Y., et al., 2011, MNRAS, 410, L13
\bibitem[Matsuda et al.(2012)]{matsuda2012}
Matsuda Y., et al., 2012, MNRAS, 425, 878
\bibitem[Miley et al.(2004)]{miley2004}
Miley G., et al., 2004, Nature, 427, 47
\bibitem[Miley \& De Breuck(2008)]{miley2008}
Miley G. \& De Breuck, 2008, ARAA, 15, 67
\bibitem[Miyazaki et al.(2002)]{miyazaki2002}
Miyazaki S., et al., 2002, PASJ, 54, 833
\bibitem[Mori \& Umemura(2006)]{mori2006}
Mori M., Umemura M., 2006, Nature, 440, 644
\bibitem[Nelson et al.(2013)]{nelson2013} Nelson, D., 
Vogelsberger, M., Genel, S., et al. 2013, MNRAS, 429, 3353 
\bibitem[Nilsson et al.(2006)]{nilsson2006}
Nilsson K. K., Fynbo J. P. U., M\o ller P., Sommer-Larsen J., Ledoux C., 
2006, A\&A, 452, L23
\bibitem[Ohyama et al.(2003)]{ohyama2003}
Ohyama Y., et al., 2003, ApJ, 591, L9
\bibitem[Oke (1974)]{oke1974}
Oke J. B., 1974, ApJS, 27, 21
\bibitem[Orsi et al.(2008)]{orsi2008}
Orsi A., lacey C., Baugh C. M., Infante L., 2008, MNRAS, 391, 1589
\bibitem[Ouchi et al.(2004)]{ouchi2004}
Ouchi M., et al., 2004, ApJ, 611, 660 
\bibitem[Ouchi et al.(2005)]{ouchi2005}
Ouchi M., et al., 2005, ApJ, 620, L1
\bibitem[Ouchi et al.(2008)]{ouchi2008}
Ouchi M., et al., 2008, ApJS, 176, 301
\bibitem[Ouchi et al.(2010)]{ouchi2010}
Ouchi M., et al., 2010, ApJ, 723, 869
\bibitem[Overzier et al.(2008)]{overzier2008}
Overzier R. A., et al. 2008, ApJ, 673, 143
\bibitem[Overzier et al.(2013)]{overzier2013}
Overzier R. A., Nesvadba N. P. H., Dijkstra M., Hatch N. A., Lehnert M. D., 
Villar-Martin\i\' n M., Wilman R. J., Zirm A. W., ApJ, 771, 89
\bibitem[Pentericci et al.(2002)]{pentericci2002}
Pentericci L., Kurk J. D., Carilli C. L., Harris D. E., Miley G. K., 
Ro\" ttgering H. J. A., 2002, A\&A, 396, 109
Moorwood A. F. M., Adelberger K. L., Giavalisco M., 2001, ApJ, 554, 981
\bibitem[Rees \& Ostriker(1977)]{rees1977}
Rees M. J., Ostriker J. P., 1977, MNRAS, 179, 541
\bibitem[Reuland et al.(2003)]{reuland2003}
Reuland M., et al., 2003, ApJ, 592, 755
\bibitem[Rhoads et al.(2000)]{rhoads2000}
Rhoads J.~E., Malhotra S., Dey A., Stern D., Spinrad H., Jannuzi B.~T., 
2000, ApJ, 545, L85
\bibitem[R\" ottgering et al.(1995)]{rottgering1995}
R\" ottgering H. J. A., Hunstead R., Miley G. K., van Ojik R>, 
Wieringa M. H., 1995, MNRAS, 277, 389
\bibitem[Saito et al.(2006)]{saito2006}
Saito T., Shimasaku K., Okamura S., Ouchi M., Akiyama M., Yoshida M., 
2006, ApJ, 648, 54
\bibitem[Saito et al.(2008)]{saito2008}
Saito T., Shimasaku K., Okamura S., Ouchi M., Akiyama M., Yoshida M. Ueda Y., 
2008, ApJ, 675, 1076
\bibitem[Schlegel, Finkbeiner, \& Davis (1998)]{schlegel1998} 
Schlegel D.~J., Finkbeiner D.~P., Davis M., 1998, ApJ, 500, 525
\bibitem[Smail \& Blundell (2013)]{smail2013}
Smail I., Blundell M., 2013, MNRAS, 434, 3246
\bibitem[Smith \& Jarvis (2007)]{smith2007}
Smith D. J. B., Jarvis M. J., 2007, MNRAS, 378, L49
\bibitem[Springel et al.(2005)]{springel2005}
Springel V. et al., 2005, Nature, 435, 2
\bibitem[Stone(1996)]{stone1996} 
Stone R.~P.~S., 1996, ApJS, 107, 423 
\bibitem[Steidel et al.(2005)]{steidel2005}
Steidel C. C., Adelberger K. L., Shapley A. E., Erb D. K., Reddy N. A., 
Pettini M., 2005, ApJ, 626, 44
\bibitem[Steidel et al.(2010)]{steidel2010}
Steidel C. C., Erb D. K., Shapley A. E., Pettini M., Reddy N., 
Bogosalievi\' c M., Rudie G. C., Rakic O., 2010, ApJ, 717, 289
\bibitem[Taniguchi (2004)]{taniguchi2004}
Taniguchi Y., 2004, Proc. Japan-German Seminar, Sendai 2001 
(Eds: Arimoto N. and Duschi W.), 107
\bibitem[Taniguchi \& Shioya (2000)]{taniguchi2000} 
Taniguchi Y., Shioya Y., 2000, ApJ, L13
\bibitem[Tenorio-Tagle et al.(1999)]{tenoriotagle1999} 
Tenorio-Tagle G., Silich S. A., Kunth D., Terlevich E., Terlevich R., 
1999, MNRAS, 309, 332
\bibitem[Uchimoto et al.(2008)]{uchimoto2008}
Uchimoto Y. K., et al., 2008, PASJ, 60, 683
\bibitem[Uchimoto et al.(2012)]{uchimoto2012}
Uchimoto Y. K., et al., 2012, ApJ, 750, 116
\bibitem[van der Wel et al. (2011)]{vanderwel2011}
van der Wel A., et al., 2011, ApJ, 742, 111
\bibitem[van Ojik et al.(1996)]{vanojik1996}
van Ojik R., R\" ottgering H. J. A., Carilli C., Bremer M. N., 
Macchetto F. A., 1996, A\&A, 313, 25
\bibitem[van Ojik et al.(1997)]{vanojik1997}
van Ojik R., R\" ottgering H. J. A., Miley G. K., Hunstead R. W., 1997, 
A\&A, 317, 358
\bibitem[Veilleux et al.(2005)]{veilleux2005}
Veilleux S., Cecil G. Bland-Hawthorn J., 2005, ARAA, 43, 769
\bibitem[Venemans et al.(2002)]{venemans2002}
Venemans B., et al., 2002, ApJ, 569, L11
\bibitem[Venemans et al.(2007)]{venemans2007}
Venemans B., et al., 2007, A\&A, 461, 823
\bibitem[Villar-Martin et al.(2007)]{villar-martin2007}
Villar-Martin M., S\' anchez S. F., Humphrey A., Dijkstra M., 
di Serego Alighieri S., De Breuck C., Gonzalez Delgado R., 2007, 
MNRAS, 378, 416
\bibitem[Wilman et al.(2005)]{wilman2005}
Wilman R. J., Gerssen J., Bower R. G., Morris S. L., Bacon R., 
de Zeeuw P. T., Davies R. L., 2005, Nature, 436, 227
\bibitem[Yagi et al.(2002)]{yagi2002}
Yagi M., Kashikawa N., Sekiguchi M., Doi M., Yasuda N., Shimasaku K., 
Okamura S., 2002, AJ, 123, 66 
\bibitem[Zirm et al.(2005)]{zirm2005}
Zirm A. W., et al., 2005, ApJ, 630, 68
\end{thebibliography}
\end{document}